\pdfoutput=1

\documentclass[11pt,twoside,a4paper,cmspaper,final,collab]{cms-tdr}
\def\svnVersion{6ce6d12-D}\def\svnDate{2020/01/23}\def\cmsCernNoTag{CERN-EP-2019-163}\def\cmsCernDate{\today}\def\cmsMessage{"Published in the European Physical Journal C as \href{http://dx.doi.org/10.1140/epjc/s10052-019-7593-7}{\doi{10.1140/epjc/s10052-019-7593-7}.}"}

\begin{document}\cmsNoteHeader{TOP-18-003}

\hyphenation{had-ron-i-za-tion}
\hyphenation{cal-or-i-me-ter}
\hyphenation{de-vices}
\newlength\cmsTabSkip\setlength{\cmsTabSkip}{1.5ex}
\newlength\cmsFigWidth
\ifthenelse{\boolean{cms@external}}{\setlength\cmsFigWidth{0.49\textwidth}}{\setlength\cmsFigWidth{0.80\textwidth}} 
\newcommand{\sslumi}{137\fbinv}
\newcommand{\Njets}{\ensuremath{N_\text{jets}}\xspace}
\newcommand{\Nbjets}{\ensuremath{N_\PQb}\xspace}
\newcommand{\Nleps}{\ensuremath{N_\ell}\xspace}
\newcommand{\tttt}{\ensuremath{\ttbar\ttbar}\xspace}
\newcommand{\ttZ}{\ensuremath{\ttbar\cPZ}\xspace}
\newcommand{\ttW}{\ensuremath{\ttbar\PW}\xspace}
\newcommand{\ttH}{\ensuremath{\ttbar\PH}\xspace}
\newcommand{\xsectttt}{\ensuremath{\sigma(\Pp\Pp\to\tttt})\xspace}
\newcommand{\POWHEGBOX} {{\textsc{powheg box}}\xspace}

\cmsNoteHeader{TOP-18-003}
\title{Search for production of four top quarks in final states with same-sign or multiple leptons in proton-proton collisions at $\sqrt{s}=13$\TeV}
\titlerunning{Search for production of four top quarks at 13\TeV}

\author*[inst1]{Giovanni Zevi Della Porta}

\date{\today}

\abstract{
The standard model (SM) production of four top quarks
(\tttt) in proton-proton collisions is studied by the CMS Collaboration.
The data sample, collected during the 2016--2018 data taking of the LHC,  corresponds to
an integrated luminosity of 137\fbinv at a center-of-mass energy of 13\TeV.
The events are required to contain two
same-sign charged leptons (electrons or muons) or at least three leptons, and jets.
The observed and expected significances for the \tttt signal
are respectively 2.6 and 2.7 standard deviations, and the
\tttt cross section is measured to be
$12.6^{+5.8}_{-5.2}\unit{fb}$. The results are used to constrain the
Yukawa coupling of the top quark to the Higgs boson, $y_{\PQt}$, yielding a
limit of $\abs{y_{\PQt}/y_{\PQt}^{\mathrm{SM}}} < 1.7$ at $95\%$ confidence level, where
$y_{\PQt}^{\mathrm{SM}}$ is the SM value of $y_{\PQt}$. They are also used to constrain
the oblique parameter of the Higgs boson in an effective field theory framework, $\hat{H}<0.12$.
Limits are set on the production of a heavy scalar or pseudoscalar boson in Type-II two-Higgs-doublet  and simplified
dark matter models, with exclusion limits reaching 350--470\GeV and
350--550\GeV for scalar and pseudoscalar bosons, respectively.
Upper bounds are also set on couplings of the top quark to new light particles.
}

\hypersetup{%
pdfauthor={CMS Collaboration},%
pdftitle={Search for production of four top quarks in final states with same-sign or multiple leptons in proton-proton collisions at sqrt(s)=13 TeV},%
pdfsubject={CMS},%
pdfkeywords={CMS, physics, top quarks}}

\maketitle

\section{Introduction}
\label{sec:intro}

\begin{figure*}[hbtp]
\centering
\includegraphics[width=.25\textwidth]{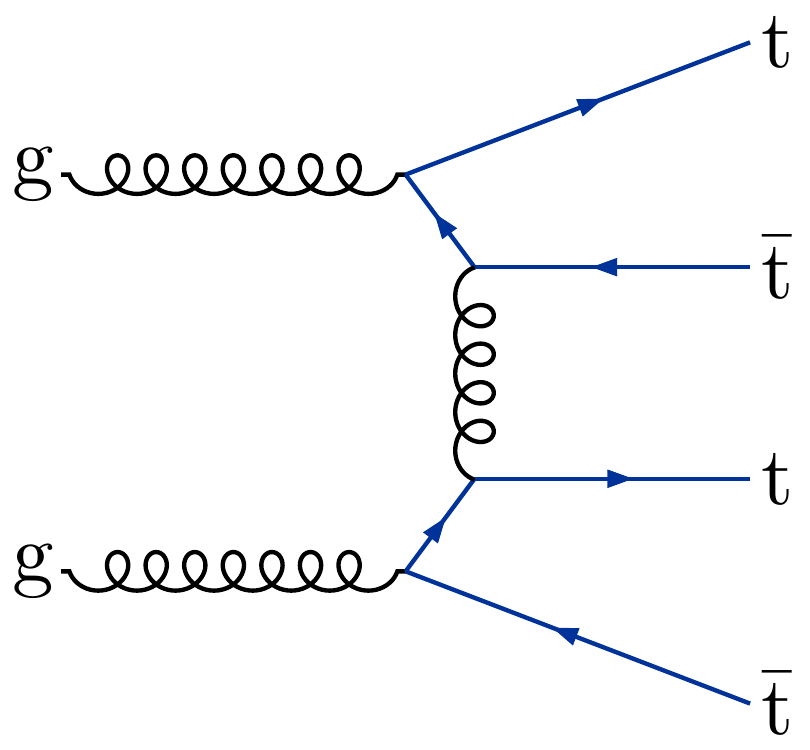}
\hspace{0.10\textwidth}
\includegraphics[width=.25\textwidth]{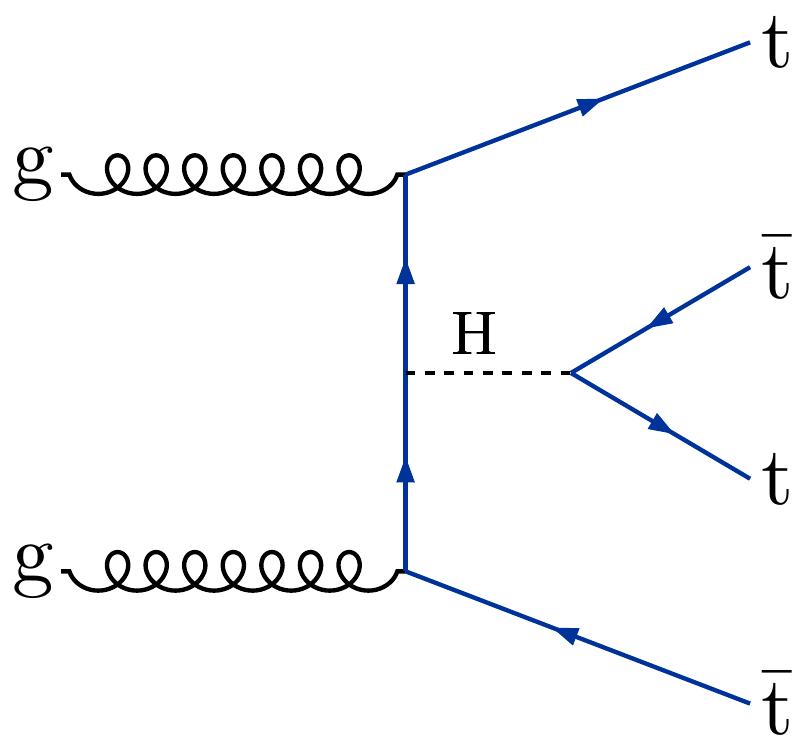}
\caption{Typical Feynman diagrams for \tttt production at leading order in the SM.}
\label{fig:feyndiagrams}
\end{figure*}

The production of four top quarks (\tttt) is a rare standard model (SM) process, with a predicted cross section
of $\xsectttt = 12.0^{+2.2}_{-2.5}\unit{fb}$ in proton-proton ($\Pp\Pp$) collisions at a center-of-mass energy of $13$\TeV,
as calculated at next-to-leading-order (NLO) accuracy for both quantum chromodynamics and electroweak interactions~\cite{Frederix:2017wme}.
Representative leading-order (LO) Feynman diagrams for SM production of \tttt are shown in Fig.~\ref{fig:feyndiagrams}.

The \tttt cross section can be used to constrain the magnitude and CP properties of the Yukawa coupling of the top quark to the Higgs boson~\cite{TopYukawaTTTT, TopYukawaTTTTnew}.
Moreover, \tttt production can be significantly enhanced by beyond-the-SM (BSM) particles and interactions.
New particles coupled to the top quark,
such as heavy scalar and pseudoscalar bosons predicted in Type-II two-Higgs-doublet models (2HDM)~\cite{Dicus:1994bm,Craig:2015jba,Craig:2016ygr}
and by simplified models of dark matter (DM)~\cite{Boveia:2016mrp, Albert:2017onk},
can contribute to \xsectttt when their masses are larger than twice the mass of the top quark, with diagrams similar to Fig.~\ref{fig:feyndiagrams} (right).
Additionally, less massive particles can enhance \xsectttt via off-shell contributions~\cite{Alvarez:2016nrz}.
In the model-independent framework of SM effective field theory,  four-fermion couplings~\cite{Hartland:2019bjb},
as well as a modifier to the Higgs boson propagator~\cite{ObliqueHiggs2019}, can be constrained through a measurement of \xsectttt.
Conversely, models with new particles with masses on the order of 1$\TeV$, such as gluino pair production in the
framework of supersymmetry~\cite{Ramond:1971gb,Golfand:1971iw,Neveu:1971rx,Volkov:1972jx,Wess:1973kz,Wess:1974tw,Fayet:1974pd,Nilles:1983ge,Martin:1997ns,Farrar:1978xj},
are more effectively probed through studies of \tttt production in boosted events or by requiring very large imbalances in momentum.

Each top quark primarily decays to a bottom quark and a \PW\ boson, and each \PW\ boson decays to either leptons or quarks. As a result, the \tttt final state contains jets mainly
from the hadronization of light (\PQu, \PQd, \PQs, \PQc) quarks (light-flavor jets) and \PQb~quarks (\PQb~jets),
and can also contain isolated charged leptons and missing transverse momentum arising from emitted neutrinos.
Final states with either two same-sign leptons or at least three leptons, considering $\PW \to \ell\nu$ ($\ell = \Pe$ or $\mu$) and including leptonic decays of $\tau$ leptons, correspond to a combined branching fraction of approximately 12\%~\cite{PDG}.
The relatively low levels of background make these channels the most sensitive to \tttt events produced with SM-like kinematic properties~\cite{SLOS2016}.

Previous searches for \tttt production in 13\TeV $\Pp\Pp$~collisions
were performed by the ATLAS~\cite{ATLAStttt2016ss, ATLAStttt2016} and CMS~\cite{SS2016,tttt2016,SLOS2016} Collaborations.
The most sensitive results, based on an integrated luminosity of approximately $36\fbinv$ collected by each experiment,
 led to cross section measurements of $28.5^{+12}_{-11}\unit{fb}$ with an observed (expected) significance of  $2.8~(1.0)$ standard deviations by ATLAS~\cite{ATLAStttt2016},
 and $13^{+11}_{-9}\unit{fb}$ with an observed (expected) significance of $1.4~(1.1)$ standard deviations by CMS~\cite{SLOS2016},
both consistent with the SM prediction.

The analysis described in this paper improves upon the CMS search presented in Ref.~\cite{tttt2016}, and supersedes the results, by taking advantage of
upgrades to the CMS detector and by optimizing the definitions of the signal regions for the integrated luminosity of \sslumi.
The reference cross section for SM \tttt, $12.0^{+2.2}_{-2.5}\unit{fb}$, used to determine the expected statistical significance of the search, as well as in interpretations for which SM \tttt is a background,
includes NLO electroweak effects, in contrast to the $9.2^{+2.9}_{-2.4}\unit{fb}$~\cite{MADGRAPH5} used in the previous search.
In addition to the analysis strategy used
in the previous search, a new multivariate classifier is defined to
maximize the sensitivity to the SM \tttt signal.

\section{Background and signal simulation}
\label{sec:samples}

Monte Carlo (MC) simulated samples at NLO are used to evaluate the signal acceptance for the SM \tttt process and to estimate
the backgrounds from diboson ($\PW\PZ$, $\PZ\PZ$, $\PZ\Pgg$, $\PW^{\pm}\PW^{\pm}$) and triboson ($\PW\PW\PW$, $\PW\PW\PZ$, $\PW\PZ\PZ$, $\PZ\PZ\PZ$, $\PW\PW\Pgg$, $\PW\PZ\Pgg$) processes.
Simulated samples generated at NLO are also used to estimate backgrounds from associated production of single top quarks and vector bosons ($\PQt\PW\PZ$,
$\PQt\PZ\PQq$, $\PQt\Pgg$), or \ttbar produced in association with a single boson (\ttW, \ttZ, \ttH, $\ttbar \Pgg$).
Three separate sets of simulated events for each process are used in order to match the different data-taking conditions and algorithms used in 2016, 2017, and 2018.
Most samples are generated using the \MGvATNLO 2.2.2 (2.4.2) program~\cite{MADGRAPH5} at NLO for 2016 samples (2017 and 2018 samples) with at most two additional partons in the matrix element calculations.
In particular, the \ttW sample is generated with up to one additional parton, and \ttZ and \ttH with no additional partons.
The \ttZ sample, which includes $\ttZ/\gamma^{*}\to\ell\ell$, is generated with a dilepton invariant mass greater than $1\GeV$.
For the $\PW\PZ$ sample used with 2016 conditions, as well as all $\PZ\PZ$ and \ttH samples, the \POWHEGBOX~v2~\cite{Melia:2011tj,Nason:2013ydw} program is used.
The \MGvATNLO generator at LO with up to three additional partons, scaled to NLO cross sections,
is used to produce a subset of samples for some of the data taking periods: $\PW\Pgg$ (2016), $\ttbar \Pgg$ (2017 and 2018), $\PQt\PZ\PQq$ (2018), and $\PQt\Pgg$ (2018)~\cite{MADGRAPH5}.
Other rare backgrounds, such as \ttbar production in association with dibosons ($\ttbar\PW\PW$, $\ttbar\PW\PZ$, $\ttbar\PZ\PZ$, $\ttbar\PW\PH$,
$\ttbar\PZ\PW$, $\ttbar\PH\PH$) and triple top quark production
($\ttbar\PQt$, $\ttbar\PQt\PW$), are  generated using LO \MGvATNLO without additional partons, and scaled to NLO cross sections~\cite{LHCHXSWG4}.
The background from radiative top decays, with $\gamma^{*}\to\ell\ell$, was found to be negligible in this analysis.

The top quark associated production modes for a heavy scalar (\PH) or pseudoscalar (\PSA) in the mass range of [350, 650]\GeV,
$\ttbar\PH/\PSA$, $\PQt\PQq\PH/\PSA$, and $\PQt\PW\PH/\PSA$,
with subsequent decays of $\PH/\PSA$ into a pair of top quarks,
are  generated using LO \MGvATNLO, with one additional parton for all but the $\PQt\PQq\PH/\PSA$ production mode.
In the context of type-II 2HDM, these samples are scaled to LO cross sections obtained
with \MGvATNLO model, ``2HDMtII'' \cite{Degrande:2014vpa,Christensen:2009jx}.
For the choice $\tan\beta = 1$ in the alignment limit~\cite{Dev:2014yca}, where $\tan\beta$ represents
the ratio of vacuum expectation values of the two Higgs doublets,
these cross sections reproduce those of Ref.~\cite{Craig:2016ygr}, which were also used in the previous CMS result~\cite{tttt2016}.
In the context of simplified models of dark matter, these samples are scaled to LO cross sections obtained with the model used in Ref.~\cite{DMsingletopCMS},
which includes kinematically accessible decays of the mediator into a pair of top quarks.
The processes are simulated in the narrow-width approximation, suitable for the parameter space studied here,
in which the width of the mediator is 5\% of its mass or less.
Samples and cross sections used for constraining the modified Higgs boson propagator are generated using \MGvATNLO at LO,
matching the prescription of Ref.~\cite{ObliqueHiggs2019}.
Cross sections used for SM \tttt enhanced by scalar and vector off-shell diagrams are obtained at LO
from Ref.~\cite{Alvarez:2016nrz}.

{\tolerance=800
The NNPDF3.0LO (NNPDF3.0NLO)~\cite{Ball:2014uwa} parton distribution
functions (PDFs) are used to generate all LO (NLO) 2016 samples, while NNPDF3.1 next-to-next-to-leading order~\cite{Ball:2017nwa} is used for 2017 and 2018 samples.
Parton showering and hadronization,  as well as $\PW^{\pm}\PW^{\pm}$ production from double-parton scattering, are modeled by the
\PYTHIA~8.205~\cite{Sjostrand:2007gs} program for 2016 samples and \PYTHIA~8.230~\cite{Sjostrand:2014zea} for 2017 and 2018 samples, while
the MLM~\cite{Alwall:2007fs} and FxFx~\cite{Frederix:2012ps} prescriptions are employed in matching
additional partons from the matrix element calculations to those from parton showers for the LO and NLO samples, respectively.
The underlying event modeling uses the CUETP8M1 tune~\cite{Skands:2014pea,Khachatryan:2015pea} for 2016, and CP5~\cite{Sirunyan:2019dfx}
for 2017 and 2018 data sets, respectively.
The top quark mass in the Monte Carlo programs is set to $172.5\GeV$.  The \GEANTfour package~\cite{Geant} is used to model the response of the CMS detector.
Additional $\Pp\Pp$ interactions (pileup) within the same or nearby bunch crossings are also included in the simulated events.
\par}

\section{ The CMS detector and event reconstruction}
\label{sec:cms}

The central feature of the CMS detector is a superconducting solenoid of 6\unit{m} internal diameter, providing a magnetic field of 3.8\unit{T}. Within the solenoid volume are a silicon pixel and strip tracker, a lead tungstate crystal electromagnetic calorimeter (ECAL), and a brass and scintillator hadron calorimeter (HCAL), each composed of a barrel and two endcap sections. Forward calorimeters extend the pseudorapidity ($\eta$) coverage provided by the barrel and endcap detectors. Muons are detected in gas-ionization chambers embedded in the steel flux-return yoke outside the solenoid.
A more detailed description of the CMS detector, together with a definition of the coordinate system used and the relevant variables, can be found in Ref.~\cite{Chatrchyan:2008zzk}.

Events of interest are selected using a two-tiered trigger system~\cite{Khachatryan:2016bia}. The first level, composed of custom hardware processors, uses information from the calorimeters and muon detectors to select events at a rate of around 100\unit{kHz} within a time interval of less than 4\mus. The second level, known as the high-level trigger, consists of a farm of processors running a version of the full event reconstruction software optimized for fast processing, and reduces the event rate to around 1\unit{kHz} before data storage.

{\tolerance=800
The reconstructed vertex with the largest value of summed physics-object
squared-transverse-momentum is taken to be the primary $\Pp\Pp$ interaction vertex. The physics
objects are the jets, clustered using the jet finding
algorithm~\cite{Cacciari:2008gp,Cacciari:2011ma} with the tracks assigned to
the vertex as inputs, and the associated missing transverse momentum, taken as
the negative vector sum of the transverse momentum (\pt) of those jets.
\par}

The particle-flow algorithm~\cite{CMS-PRF-14-001} aims to reconstruct and
identify each individual particle in an event, with an optimized combination of
information from the various elements of the CMS detector.  The energy of
photons is directly obtained from the ECAL measurement. The energy of electrons
is determined from a combination of the electron momentum at the primary
interaction vertex as determined by the tracker, the energy of the
corresponding ECAL cluster, and the energy sum of all bremsstrahlung photons
spatially compatible with the electron track~\cite{Khachatryan:2015hwa}. The
momentum of muons is obtained from the curvature of the corresponding track,
combining information from the silicon tracker and the muon
system~\cite{Chatrchyan:2012xi}. The energy of charged hadrons is determined
from a combination of their momentum measured in the tracker and the matching
ECAL and HCAL energy deposits, corrected for the response function of the
calorimeters to hadronic showers. The energy of neutral hadrons is obtained
from the corresponding corrected ECAL and HCAL energies.

Hadronic jets are clustered from neutral PF candidates and charged PF candidates associated with the primary vertex, using the anti-\kt algorithm~\cite{Cacciari:2008gp, Cacciari:2011ma} with a distance parameter
of 0.4.
The jet momentum is determined as the vectorial sum of all PF candidate momenta in the jet. An offset correction is applied to jet energies to take into account the contribution from pileup~\cite{cacciari-2008-659}.
Jet energy corrections are derived from simulation and are improved with in situ measurements of the energy balance in dijet, multijet, $\gamma$+jet, and leptonically decaying Z+jet events~\cite{Khachatryan:2016kdb, CMS-PAS-JME-16-003}.
Additional selection criteria are applied to each jet to remove jets potentially affected by instrumental effects or reconstruction failures~\cite{CMS-PAS-JME-16-004}.
Jets originating from \PQb~quarks are identified as \PQb-tagged jets using a deep neural network algorithm, DeepCSV~\cite{Sirunyan:2017ezt},
with a working point chosen such that the efficiency to identify a \PQb~jet is  55--70\% for a jet \pt between 20 and 400\GeV.
The misidentification rate is approximately 1--2\% for light-flavor and gluon jets and 10--15\% for charm jets, in the same jet \pt range.
The vector \ptvecmiss is defined as the projection on the plane perpendicular to the beams of the negative vector sum of the momenta of all reconstructed PF candidates in an event~\cite{Sirunyan:2019kia}. Its magnitude, called missing transverse momentum, is referred to as \ptmiss.

\section{Event selection and search strategy}
\label{sec:selection}

The identification, isolation, and impact parameter requirement with respect to the primary vertex, imposed on electrons and muons are the same as those of Ref.~\cite{tttt2016}
when analyzing the 2016 data set, while for the 2017 and 2018 data sets the identification of electrons and the isolation of both electrons and muons
are modified to take into account the increased pileup.
For electrons, identification is based on a multivariate discriminant using shower shape and track quality variables, while
muon identification is based on the quality of the geometrical matching between measurements in the tracker and the muon system.
The isolation requirement, introduced in Ref.~\cite{SUS-15-008}, is designed to distinguish the charged leptons produced in \PW\ and \PZ\ decays (``prompt leptons'') from
the leptons produced in hadron decays or in conversions of photons in jets, as well as hadrons misidentified as leptons (collectively defined as ``nonprompt leptons'').
The requirements to minimize charge misassignment are the same as in Ref.~\cite{tttt2016}: muon tracks are required to have a
small uncertainty in \pt and electron tracks are required to have the same charge as that obtained from
comparing a linear projection of the pixel detector hits to the position of the calorimeter deposit.
The combined efficiency to reconstruct and identify leptons is in the range of
45--80 (70--90)\% for electrons (muons),
increasing as a function of \pt and reaching the maximum value for $\pt>60\GeV$.

For the purpose of counting leptons and jets, the following requirements are applied: the number of leptons (\Nleps)
is defined to be the multiplicity of electrons and muons with $\pt > 20\GeV$ and either $\abs{\eta} < 2.5$ (electrons) or $\abs{\eta} < 2.4$ (muons), the number of jets (\Njets)
counts all jets with $\pt > 40\GeV$ and $\abs{\eta} < 2.4$, and the number of \PQb-tagged jets (\Nbjets) counts \PQb-tagged jets
with $\pt > 25\GeV$ and $\abs{\eta} < 2.4$.
In order to be included in \Njets, \Nbjets, and the \HT variable, which is defined as the scalar \pt sum of all jets in an event,
jets and \PQb-tagged jets must have an angular separation $\Delta R > 0.4$ with respect to all selected leptons.
This angular separation is defined as $\Delta R = \sqrt{\smash[b]{(\Delta\eta)^2+(\Delta\phi)^2}}$,
where $\Delta\eta$ and $\Delta\phi$ are the differences in pseudorapidity and azimuthal angle, respectively, between the directions of the lepton and the jet.

{\tolerance=800
Events were recorded using either a dilepton$+$\HT (2016) or a set of dilepton triggers (2017 and 2018).
The dilepton$+$\HT trigger requires two leptons with $\pt > 8\GeV$ and a minimum $\HT$ requirement that is fully efficient with respect to the offline requirement of $300\GeV$.
The dilepton triggers require either two muons with $\pt > 17$ and $8\GeV$, two electrons with $\pt > 23$ and $12\GeV$, or an
$\Pe\Pgm$ pair with $\pt > 23\GeV$ for the higher-$\pt$ (leading) lepton and $\pt > 12\ (8)\GeV$ for the lower-$\pt$ (trailing) electron (muon).
The trigger efficiency within the detector acceptance is measured in data to be greater than $90\%$ for $\Pe\Pe$, $\Pe\Pgm$, and $\Pgm\Pgm$ events,
and nearly $100\%$ for events with at least three leptons.
\par}

We define a baseline selection that requires $\HT > 300\GeV$ and $\ptmiss>50\GeV$,
two or more jets ($\Njets \geq 2$) and \PQb-tagged jets ($\Nbjets \geq 2$), a leading lepton with $\pt > 25\GeV$,
and a trailing lepton of the same charge with $\pt > 20\GeV$.
Events with same-sign electron pairs with an invariant mass below 12\GeV are rejected to reduce the background from
production of low-mass resonances with a charge-misidentified electron.
Events where a third lepton with $\pt>7$ (5)\GeV for electrons (muons) forms an opposite-sign (OS) same-flavor
pair with an invariant mass below 12\GeV or between 76 and 106\GeV are also rejected.
Inverting this resonance veto, the latter events are used  to populate a \ttZ background control region (CRZ) if the invariant mass is between 76 and 106\GeV and the third lepton has $\pt > 20\GeV$.
After this baseline selection, the signal acceptance is approximately 1.5\%,
including branching fractions.

Events passing the baseline selection are split into several signal and control regions,
following two independent approaches.
In the first analysis, similarly to Ref.~\cite{tttt2016} and referred to as ``cut-based'', the variables \Njets, \Nbjets, and \Nleps are used to subdivide events into 14 mutually exclusive signal regions (SRs)
and a control region (CR) enriched in \ttW background (CRW), to complement the CRZ defined above, as detailed in Table~\ref{tab:SRDefNew}.
In the boosted decision tree (BDT) analysis, the CRZ is the only control region, and the remaining events are subdivided into 17 SRs
by discretizing the discriminant output of a BDT trained to separate \tttt events from the sum of the SM backgrounds.

The BDT classifier utilizes a gradient boosting algorithm to train 500 trees with a depth of 4 using simulation, and is based on the following 19 variables:
\Njets, \Nbjets, \Nleps, \ptmiss, \HT, two alternative definitions of \Nbjets based on \PQb~tagging working points tighter or looser than the default one, the scalar
\pt sum of \PQb-tagged jets, the \pt of the three leading leptons, of the leading jet and of the sixth, seventh, and eighth jets, the azimuthal angle between the two leading leptons, the invariant mass
formed by the leading lepton and the leading jet, the charge of the leading lepton,
and the highest ratio of the jet mass to the jet \pt in the event (to provide sensitivity to boosted, hadronically-decaying
top quarks and W bosons).
Three of the most performant input variables, \Njets, \Nbjets, and \Nleps, correspond to the variables used for the cut-based analysis.
Top quark tagging algorithms to identify hadronically decaying top quarks
based on invariant masses of jet combinations, similarly to Ref.~\cite{SLOS2016}, were also tested, but did not improve the expected sensitivity.
Such algorithms could only contribute in the handful of events where all the top quark decay products were found, and these events already have very small background yields.
In each analysis, the observed and predicted yields in the CRs and SRs are used in a maximum likelihood fit with nuisance parameters to measure \xsectttt,
following the procedure described in Section~\ref{sec:results}.

\begin{table}[htb!]
\centering
\topcaption{\label{tab:SRDefNew} Definition of the 14 SRs and two CRs for the cut-based analysis.}
\begin{tabular}{c|c|c|c}
\hline
$\Nleps$                             & $\Nbjets$                 & $\Njets$ & Region \\ \hline
\multirow{9}{*}{2}                  & \multirow{4}{*}{2}        & $\leq$5  & CRW \\
                                     &                           & 6        & SR1 \\
                                     &                           & 7        & SR2 \\
                                     &                           & $\geq$8  & SR3 \\[\cmsTabSkip]
                                     & \multirow{4}{*}{3}        & 5        & SR4 \\
                                     &                           & 6        & SR5 \\
                                     &                           & 7        & SR6 \\
                                     &                           & $\geq$8  & SR7 \\[\cmsTabSkip]
                                     & \multirow{1}{*}{$\geq 4$ } & $\geq$5  & SR8 \\\hline
\multirow{6}{*}{$\geq 3$ }            & \multirow{3}{*}{2}        & 5        & SR9 \\
                                     &                           & 6        & SR10 \\
                                     &                           & $\geq$7  & SR11 \\[\cmsTabSkip]
                                     & \multirow{3}{*}{$\geq 3$ } & 4        & SR12 \\
                                     &                           & 5        & SR13 \\
                                     &                           & $\geq$6  & SR14 \\ \hline
\multicolumn{3}{c|}{Inverted resonance veto} & CRZ \\ \hline
\end{tabular}
\end{table}

\section{Backgrounds}
\label{sec:backgrounds}

In addition to the \tttt signal, several other SM processes result in final states with same-sign dileptons or at least three leptons, and several jets and \PQb~jets.
These backgrounds primarily consist of processes where \ttbar is produced in association with additional bosons that decay to leptons,
such as \ttW, \ttZ, and \ttH (mainly in the $\PH \to \PW\PW$ channel),
as well as dilepton \ttbar events with a charge-misidentified prompt-lepton and single-lepton \ttbar events with an additional nonprompt lepton.

The prompt-lepton backgrounds, dominated by \ttW, \ttZ, and \ttH, are estimated using simulated events.
Dedicated CRs are used to constrain the normalization for \ttW (cut-based analysis) and \ttZ (cut-based and BDT analyses),
while for other processes described in the next paragraph, the normalization is based on the NLO cross sections referenced in Section~\ref{sec:samples}.

Processes with minor contributions are grouped into three categories.
The ``$\ttbar\PV\PV$'' category includes the associated production of \ttbar with a pair of bosons ($\PW$, $\PZ$,  $\PH$), dominated by $\ttbar\PW\PW$.
The ``X\Pgg'' category includes processes where a photon accompanies a vector boson, a top quark, or a top-antitop quark pair.
The photon undergoes a conversion, resulting in the identification of an electron in the final state.
The category is dominated by $\ttbar \Pgg$,
with smaller contributions from $\PW\Pgg$, $\PZ\Pgg$, and $\PQt\Pgg$.
Finally, the ``Rare'' category includes all residual processes with top quarks ($\PQt\PZ\PQq$, $\PQt\PW\PZ$, $\ttbar\PQt$, and $\ttbar\PQt\PW$)
or without them ($\PW\PZ$, $\PZ\PZ$, $\PW^{\pm}\PW^{\pm}$ from single- and double-parton scattering, and triboson production).

Since the \ttW, \ttZ, and \ttH processes constitute the largest backgrounds to \tttt production, their simulated samples are corrected wherever possible to account for
discrepancies observed between data and MC simulation.
To improve the MC modeling of the
additional jet multiplicity from initial-state radiation (ISR) and final-state radiation (FSR),
simulated \ttW and \ttZ events are reweighted based on the number of ISR or FSR jets ($N_{\text{jets}}^{\mathrm{ISR/FSR}}$).
The reweighting is based on a comparison of the light-flavor jet
multiplicity in dilepton \ttbar events in data and simulation,
where the simulation is performed with the same generator settings as those of the \ttW and \ttZ samples.
The method requires exactly two jets identified as originating from \PQb~quarks in the event and assumes that all other jets are
from ISR or FSR.
The $N_{\text{jets}}^{\mathrm{ISR/FSR}}$ reweighting factors vary within the range of
[0.77,1.46]
for
$N_{\text{jets}}^{\mathrm{ISR/FSR}}$ between 1 and 4.
This correction is not applied to \ttH ($\PH \to \PW\PW$) events, which already have additional jets from the decay of the additional $\PW$ bosons.
In addition to the ISR or FSR correction, the  \ttW, \ttZ, and \ttH simulation is corrected  to improve the modeling of the flavor of additional jets,
based on the measured ratio of the $\ttbar \bbbar$ and $\ttbar \text{jj}$ cross sections, $1.7\pm0.6$ , reported in Ref.~\cite{CMSttbb}, where j represents a generic jet.
This correction results in a 70\% increase of
events produced in association with a pair of additional \PQb~jets.
Other topologies, such as those including \PQc quarks, are negligible by comparison, and no dedicated correction is performed.

The nonprompt lepton backgrounds are estimated using the ``tight-to-loose'' ratio method~\cite{SUS-15-008}.
The tight identification (for electrons) and isolation (for both electrons and muons) requirements of the SRs
are relaxed to define a loose lepton selection, enriched in nonprompt leptons.
The efficiency, $\epsilon_{\mathrm{TL}}$, for nonprompt leptons that satisfy the loose selection to also satisfy the tight selection
is measured in a control sample of single-lepton events, as a function of lepton flavor, \pt, and $\abs{\eta}$, after subtracting the prompt-lepton contamination based on simulation.
The loose selection is chosen to ensure that $\epsilon_{\mathrm{TL}}$ remains stable across the main categories of nonprompt leptons specified in Section~\ref{sec:selection},
allowing the same $\epsilon_{\mathrm{TL}}$ to be applied to samples with different nonprompt lepton composition.
For leptons failing the
tight selection, the \pt variable is redefined as the sum of the lepton \pt and the energy in the isolation cone exceeding the isolation threshold value.
This parametrization accounts for the momentum spectrum of the parent parton (the parton that produced the nonprompt lepton),
allowing the same $\epsilon_{\mathrm{TL}}$ to be applied to samples with different parent parton momenta
with reduced bias.
To estimate the number of nonprompt leptons in each SR, a dedicated set of application regions is defined,
requiring at least one lepton to fail the tight selection while satisfying the loose one (loose-not-tight). Events in these regions are then
weighted by a factor of $\epsilon_{\mathrm{TL}} / (1-\epsilon_{\mathrm{TL}})$ for each loose-not-tight lepton.
To avoid double counting the contribution of events with multiple nonprompt leptons,
events with two loose-not-tight leptons are subtracted,
and the resulting total weight is used as a prediction of the nonprompt lepton yield.

The background resulting from charge-misidentified leptons is estimated using the charge-misidentification probability measured in simulation as a function of electron \pt and $\abs{\eta}$.
This probability ranges between $10^{-5}$ and $10^{-3}$ for electrons and is at least an order of magnitude smaller for muons.
Charge-misidentified muons are therefore considered negligible, while for electrons this probability is
applied to a CR of OS dilepton events defined for each same-sign dilepton SR.
A single correction factor, inclusive in \pt and $\abs{\eta}$, is applied to the resulting estimate to account for differences between data and simulation in this probability.
A correction factor, derived from a control sample enriched in $\PZ\to\Pep\Pem$ events with one electron or positron having a misidentified charge,
is very close to unity for the 2016 simulation, while it is approximately 1.4 for the 2017 and 2018 simulation.
Even with the larger correction factors, the charge-misidentification probability is smaller in 2017 and 2018 than in 2016, due to the upgraded pixel detector~\cite{CERN-LHCC-2012-016}.

\section{Uncertainties}
\label{sec:systematics}

Several sources of experimental and theoretical uncertainty related to signal and background processes are considered in this analysis.
They are summarized, along with their estimated correlation treatment across the 2016, 2017, and 2018 data sets, in Table~\ref{tab:systSummary}.
Most sources of uncertainties affect simulated samples, while the backgrounds obtained using control samples in data (charge-misidentified and nonprompt leptons)
have individual uncertainties described at the end of this section.

The uncertainties in the integrated luminosity are 2.5, 2.3, and 2.5\% for the 2016, 2017, and 2018 data collection periods, respectively~\cite{cmsLumi2016,cmsLumi2017,cmsLumi2018}.
Simulated events are reweighted to match the distribution of the number of pileup collisions per event in data.
This distribution is derived from the instantaneous luminosity and the inelastic cross section~\cite{Sirunyan:2018nqx},
and uncertainties in the latter are propagated to the final yields, resulting in yield variations of at most 5\%.

\begin{table*}[!hbt]
\centering
    \topcaption{
   Summary of the sources of uncertainty, their values, and their impact, defined as the relative change of the
   measurement of $\sigma(\ttbar\ttbar)$ induced by one-standard-deviation variations corresponding to each uncertainty source considered separately.
    The first group lists experimental and theoretical uncertainties
    in simulated signal and background processes.
    The second group lists normalization uncertainties in the estimated backgrounds.
    Uncertainties marked (not marked) with a $\dagger$ in the first column are treated as fully correlated (fully uncorrelated) across the three years of data taking.
    }
    \label{tab:systSummary}
\small
    \begin{tabular}{lcc}
                                                         &                  & Impact on \\
    Source                                               & Uncertainty (\%) & $\sigma(\ttbar\ttbar)$ (\%) \\
\hline
    Integrated luminosity                                & 2.3--2.5         & 2 \\
    Pileup                                               & 0--5             & 1 \\
    Trigger efficiency                                   & 2--7             & 2 \\
    Lepton selection                                     & 2--10            & 2 \\
    Jet energy scale                                     & 1--15            & 9 \\
    Jet energy resolution                                & 1--10            & 6 \\
    $\cPqb$ tagging                                      & 1--15            & 6 \\
    Size of simulated sample                             & 1--25            & $<$1 \\
    Scale and PDF variations~$\dagger$                   & 10--15           & 2 \\
    ISR/FSR (signal)~$\dagger$                           & 5--15            & 2 \\
[\cmsTabSkip]
    $\ttbar \PH$ (normalization)~$\dagger$               & 25               & 5 \\
    Rare, X\Pgg, $\ttbar\PV\PV$ (norm.)~$\dagger$   & 11--20            & $<$1 \\
    $\ttbar \PZ$, $\ttbar \PW$ (norm.)~$\dagger$         & 40               & 3--4 \\
    Charge misidentification~$\dagger$                   & 20               & $<$1 \\
    Nonprompt leptons~${}\dagger$                        & 30--60           & 3 \\
    $N_{\text{jets}}^{\mathrm{ISR/FSR}}$                    & 1--30             & 2 \\
    $\sigma({\ttbar\bbbar})/\sigma({\ttbar\mathrm{jj}})$~${}\dagger$ & 35               & 11 \\
\end{tabular}

\end{table*}

The efficiency of the trigger requirements is measured in an independent data sample selected using single-lepton triggers, with an uncertainty of 2\%.
The lepton reconstruction and identification efficiency is measured using a data sample enriched in $\PZ\to\ell\ell$ events~\cite{Khachatryan:2015hwa,Chatrchyan:2012xi},
with uncertainties of up to 5 (3)\% per electron (muon).
The tagging efficiencies for \PQb~jets and light-flavor jets are measured in dedicated data samples~\cite{Sirunyan:2017ezt}, and their uncertainties result
in variations between 1 and 15\% of the signal region yields.
In all cases, simulated events are reweighted to match the efficiencies measured in data.
The uncertainty associated with jet energy corrections results
in yield variations of 1--15\% across SRs. Uncertainties in the jet energy resolution result in 1--10\% variations~\cite{Khachatryan:2016kdb}.

As discussed in
Section~\ref{sec:backgrounds}, we correct the distribution of the number of additional jets in  \ttW and \ttZ samples, with
reweighting factors varying within the range of [0.77,1.46]. We take one half of the differences from unity
as the systematic uncertainties in these factors, since they are measured in a \ttbar sample, but are applied to different processes.
These uncertainties result in yield variations up to 8\% across SRs.
Similarly, events with additional \PQb~quarks in \ttW, \ttZ, and \ttH are scaled by a factor of $1.7\pm0.6$, based on the CMS  measurement
of the ratio of cross sections $\sigma({\ttbar\bbbar})/\sigma({\ttbar\mathrm{jj}})$~\cite{CMSttbb}. The resulting uncertainty in the yields for SRs with $\Nbjets \geq 4$,
where the effect is dominant, is up to 15\%.

For background processes, uncertainties in the normalization (number of events passing the baseline selection) and shape (distribution of events across SRs) are considered,
while for signal processes, the normalization is unconstrained, and instead, we consider the uncertainty in the acceptance (fraction of events passing the baseline selection) and shape.
For each of the Rare, X\Pgg, and $\ttbar\PV\PV$ categories, normalization uncertainties are taken from the largest theoretical cross section uncertainty
in any constituent physics process, resulting in uncertainties of 20\%, 11\%, and 11\%, respectively.
For the \ttW and \ttZ processes, we set an initial normalization uncertainty of 40\%, but then allow the maximum-likelihood fit to constrain these backgrounds further using control samples in data.
For \ttH, we assign a 25\% normalization uncertainty to reflect the signal strength,
which is the ratio between the measured cross section of \ttH and its SM expectation,
of $1.26^{+0.31}_{-0.26}$ measured by CMS~\cite{Sirunyan:2018hoz}.

The shape uncertainty resulting from variations of the renormalization and factorization scales in the event generators is smaller than 15\% for backgrounds, and 10\% for the \tttt and 2HDM signals,
while the effect of the PDFs is only 1\%.
For the \tttt and 2HDM signals, the uncertainty in the acceptance from variations of the scales is 2\%. The uncertainty in the scales
that determine ISR and FSR, derived from \tttt samples, results in up to 6 and 10\% uncertainties in signal acceptance and shape, respectively.
When considering \tttt as a background in BSM interpretations, a cross section uncertainty of 20\% (based on the prediction of $12.0^{+2.2}_{-2.5}\unit{fb}$~\cite{Frederix:2017wme}) is additionally applied to the \tttt process.

The charge-misidentified and nonprompt-lepton backgrounds are assigned an uncertainty of 20 and 30\%, respectively, where the
latter is increased to 60\% for nonprompt electrons with $\pt > 50\GeV$.
For the charge-misidentified lepton background, the uncertainty is based on the agreement observed
between the prediction and data as a function of kinematic distributions, in a data sample enriched in $\PZ\to\Pep\Pem$ events with one electron or positron having a misidentified charge.
For the nonprompt-lepton background, the uncertainty is based on the agreement observed in simulation closure tests of the ``tight-to-loose'' method using multijet, \ttbar, and \PW$\!\!+\!$ jets samples.
The contamination of prompt leptons, which is subtracted based on simulation, is below 1\% in the application region, but it can be significant in the control sample where $\epsilon_{\mathrm{TL}}$ is measured, resulting in an uncertainty up to 50\% in $\epsilon_{\mathrm{TL}}$.
The statistical uncertainty in the estimate based on control samples in data is taken into account for both backgrounds. It is negligible for the
charge-misidentified lepton background, while for the nonprompt-lepton background it can be comparable or larger than the
systematic uncertainty.

Experimental uncertainties in normalization and shape are treated as fully correlated among the SRs for all signal and background processes.
Two choices of correlation across years (uncorrelated or fully correlated) were tested for each experimental uncertainty, and their impact
on the measurement of $\sigma(\ttbar\ttbar)$ was found to be smaller than 1\%.
For simplicity, these uncertainties are then treated as uncorrelated.
Systematic uncertainties in the background estimates based on control samples in data and theoretical uncertainties in the normalization of each background process
are treated as uncorrelated between processes but fully correlated among the SRs and across the three years.
Scale and PDF uncertainties, as well as uncertainties in the number of additional \PQb~quarks, are correlated between processes, signal regions, and years.
Statistical uncertainties due to the finite number of simulated events or control region events are considered uncorrelated.

\section{Results}
\label{sec:results}

\begin{figure*}[!hbt]
\centering
\includegraphics[width=.49\textwidth]{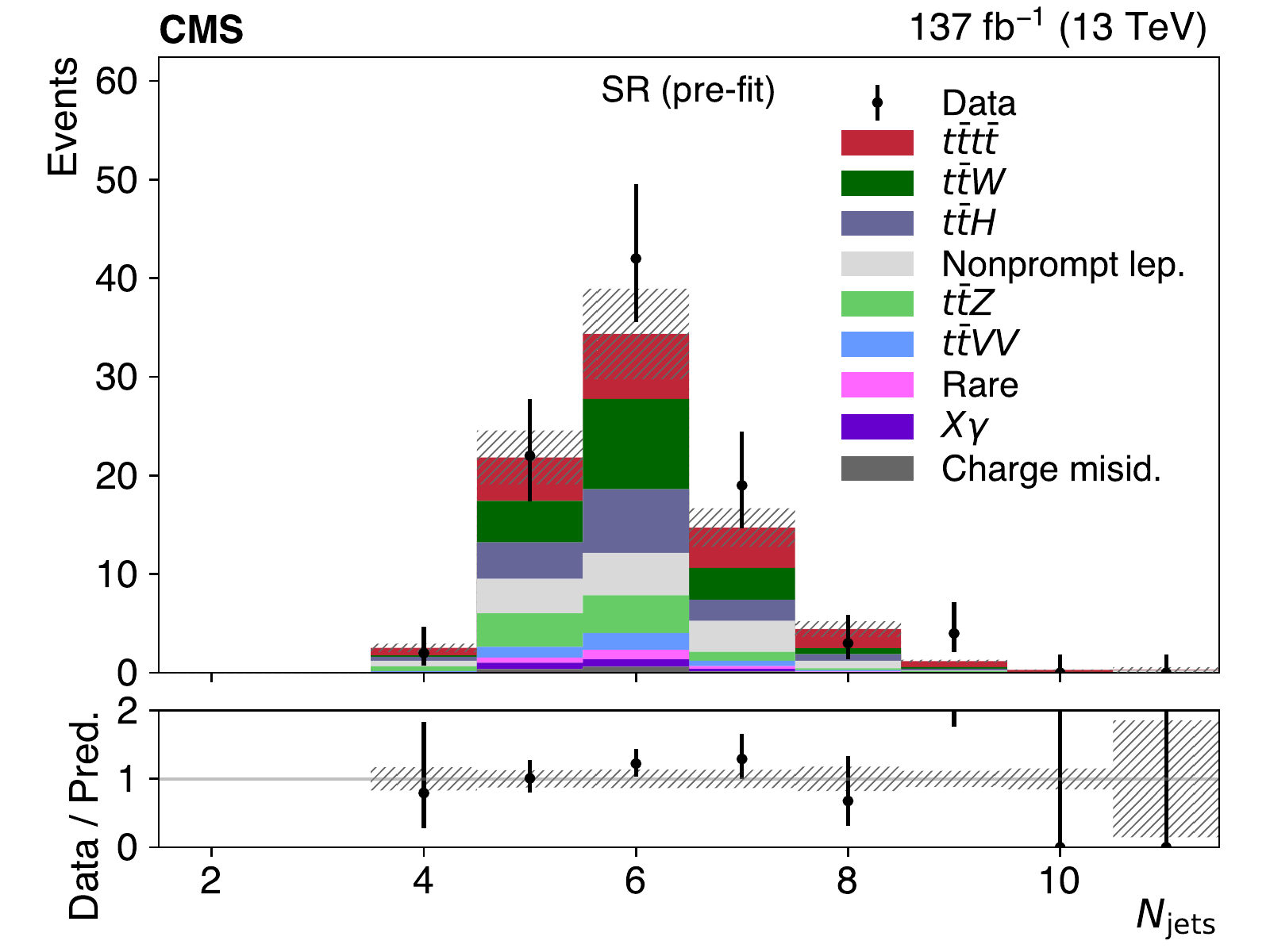}
\includegraphics[width=.49\textwidth]{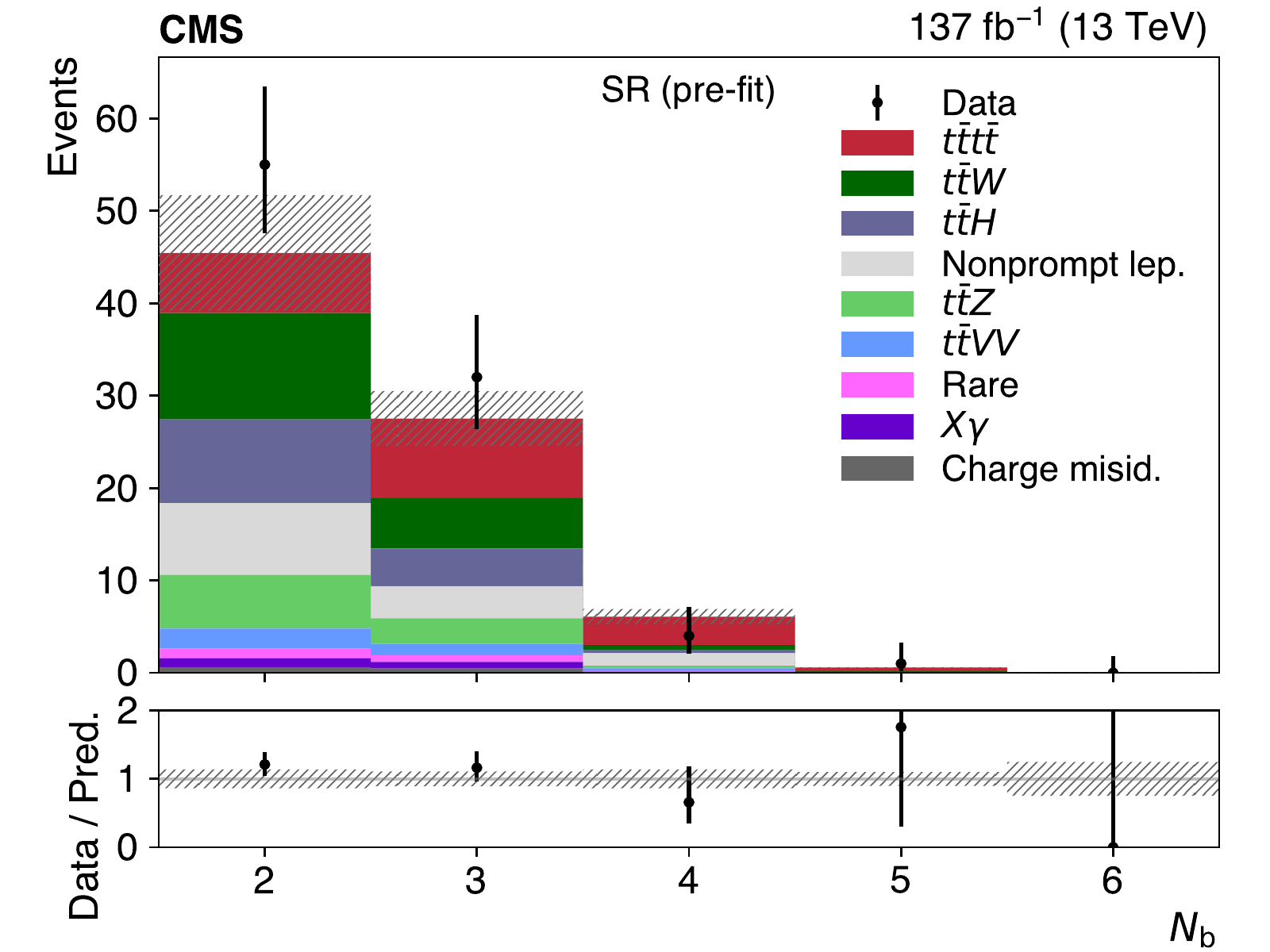}
\includegraphics[width=.49\textwidth]{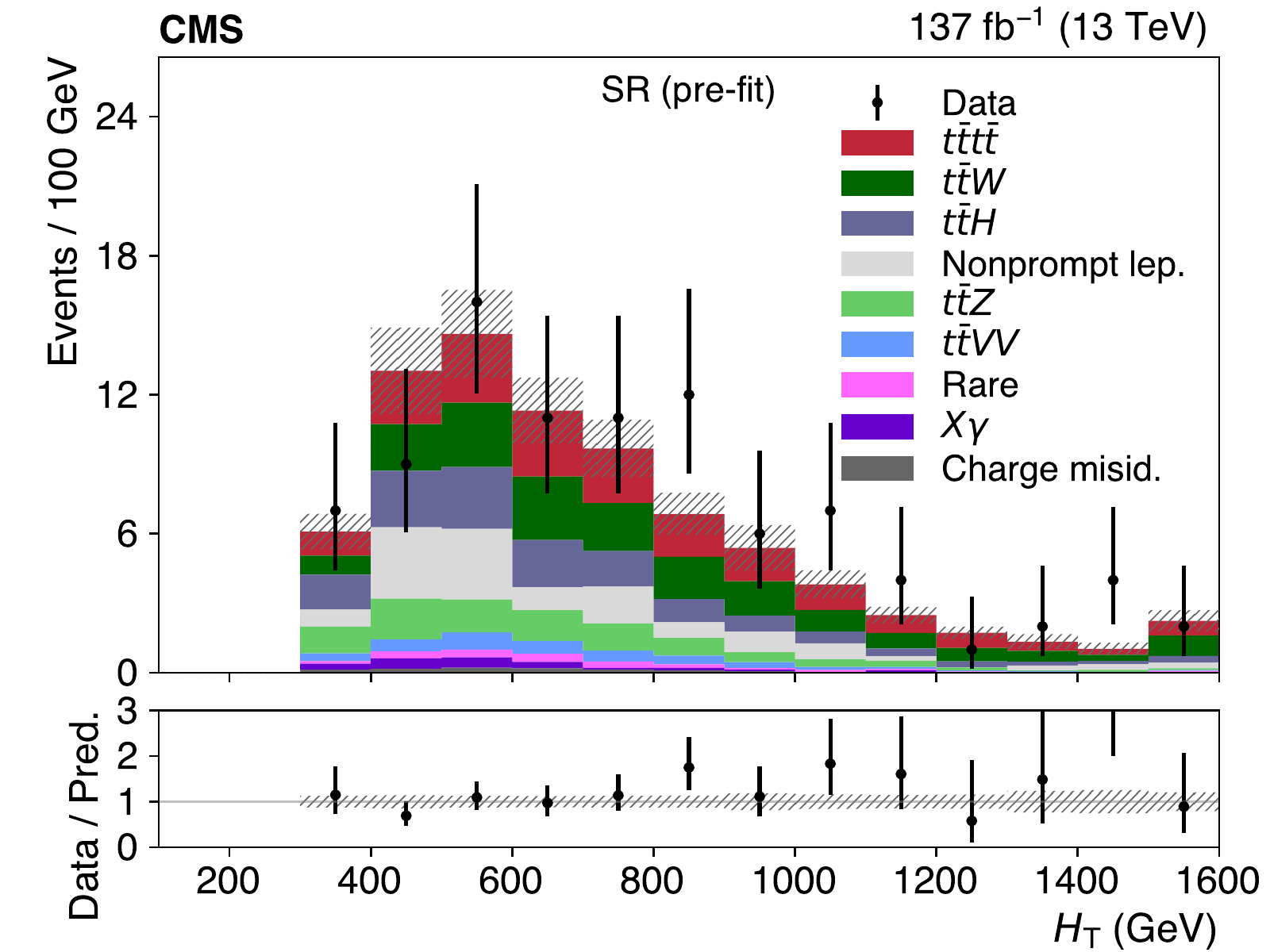}
\includegraphics[width=.49\textwidth]{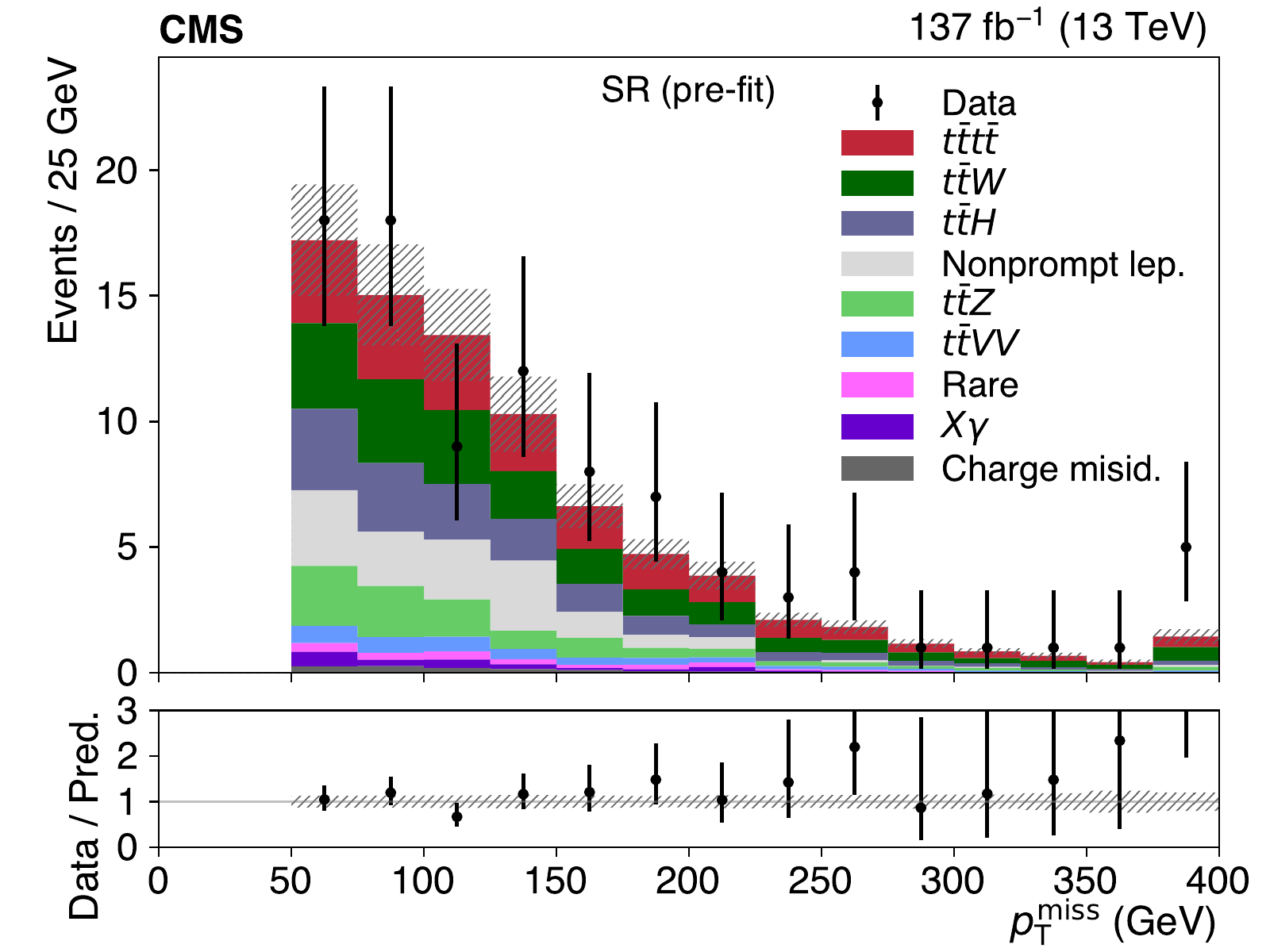}
\caption{ Distributions of \Njets (upper left), \Nbjets (upper right), \HT (lower left), and \ptmiss (lower right) in the summed SRs (1--14), before fitting to data,
where the last bins include the overflows. The hatched areas represent the total uncertainties in the SM signal and background predictions.
The \tttt signal assumes the SM cross section from Ref.~\cite{Frederix:2017wme}.
 The lower panels show the ratios of the observed event yield to the total prediction of
    signal plus background.
    }
\label{fig:kinemsr}
\end{figure*}

\begin{figure*}[!hbt]
\centering
\includegraphics[width=.49\textwidth]{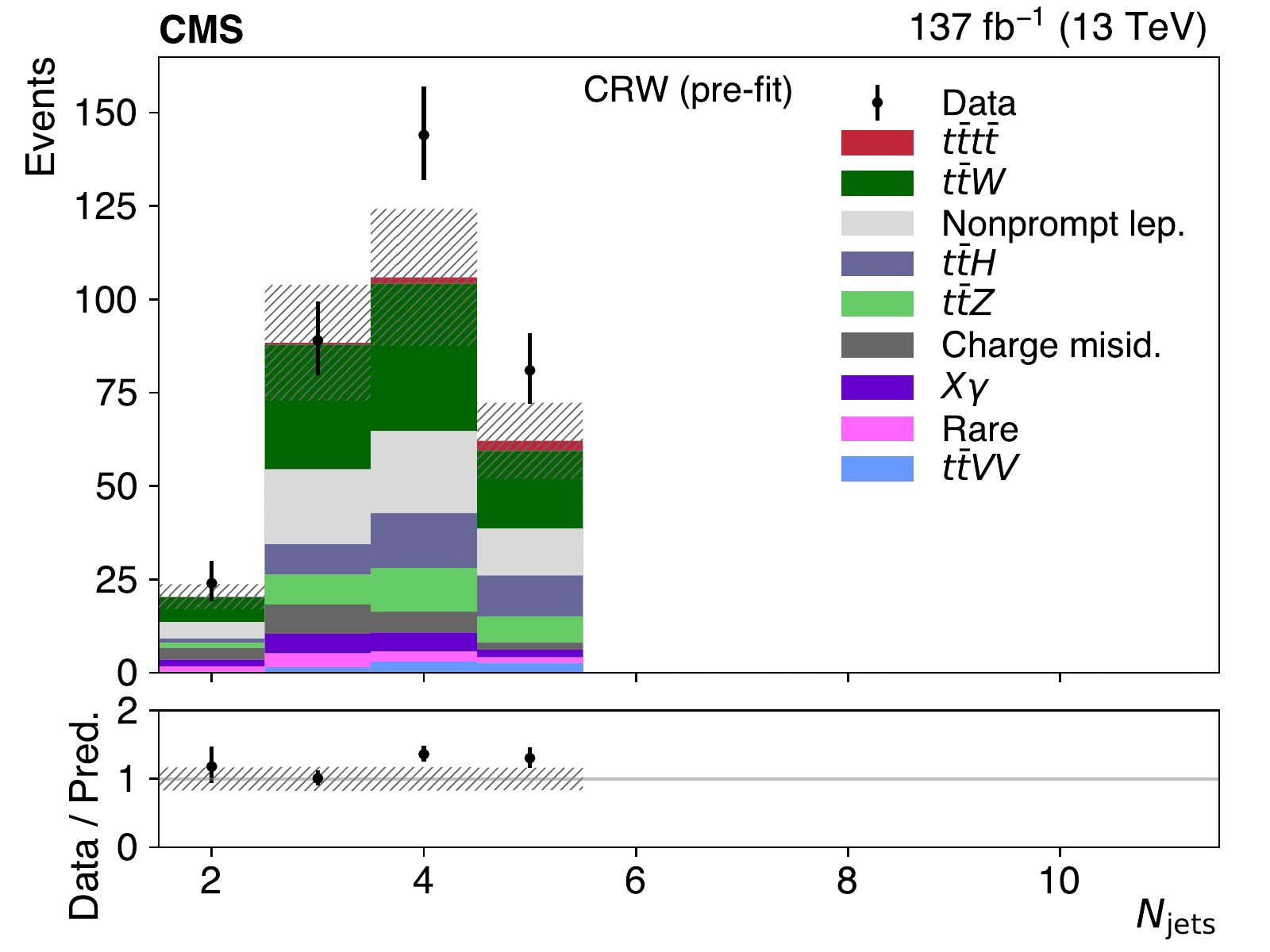}
\includegraphics[width=.49\textwidth]{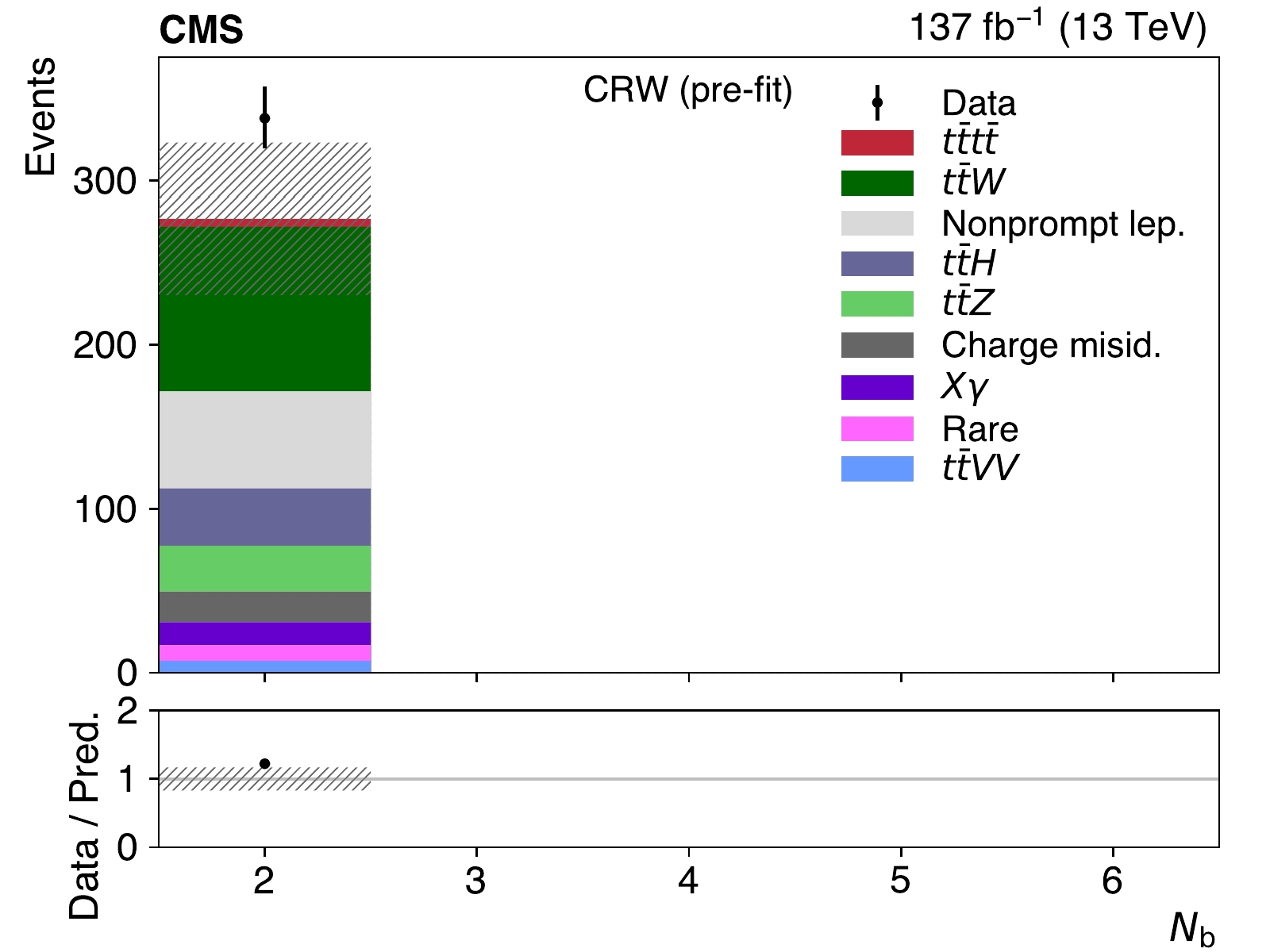}
\includegraphics[width=.49\textwidth]{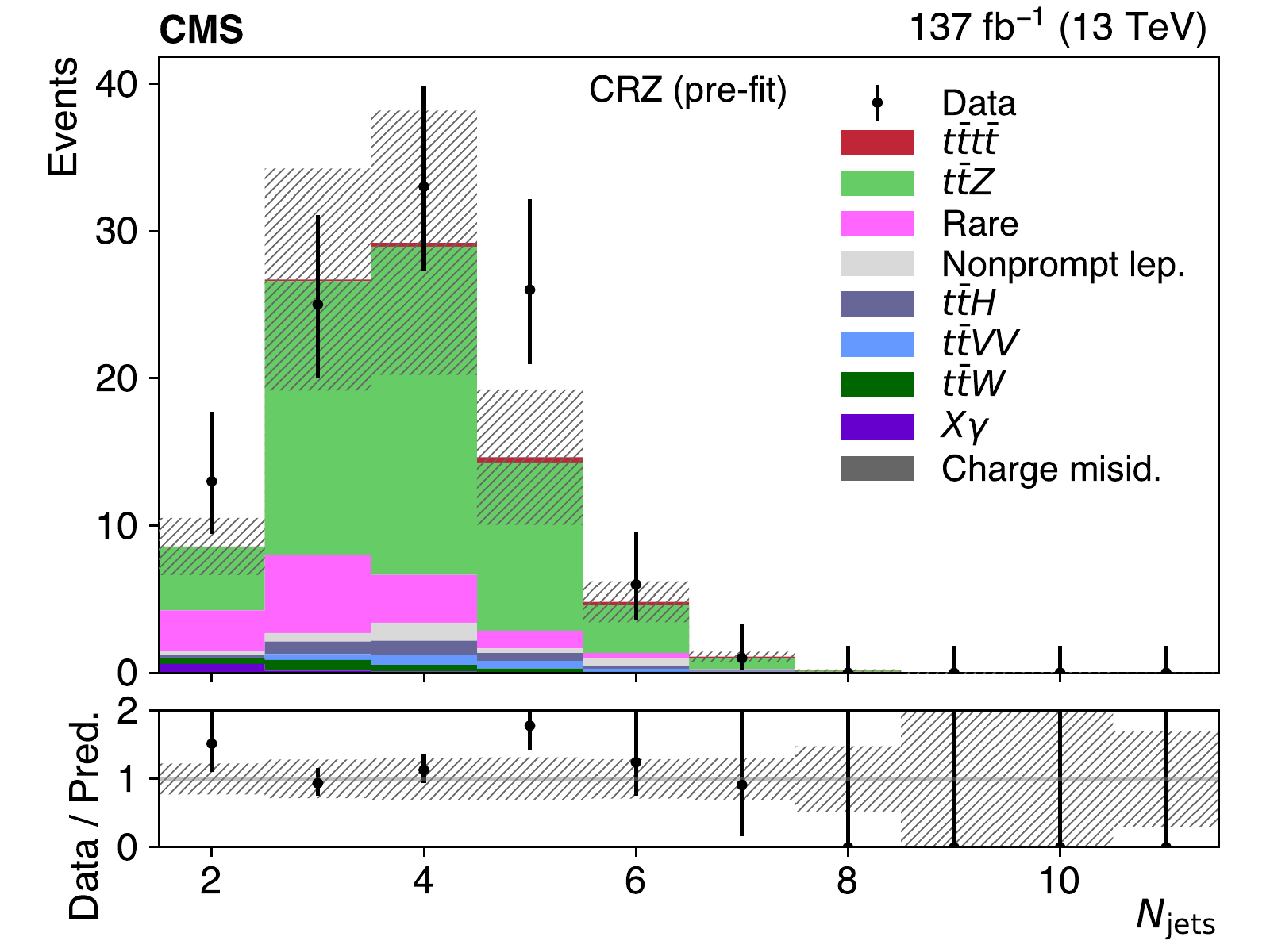}
\includegraphics[width=.49\textwidth]{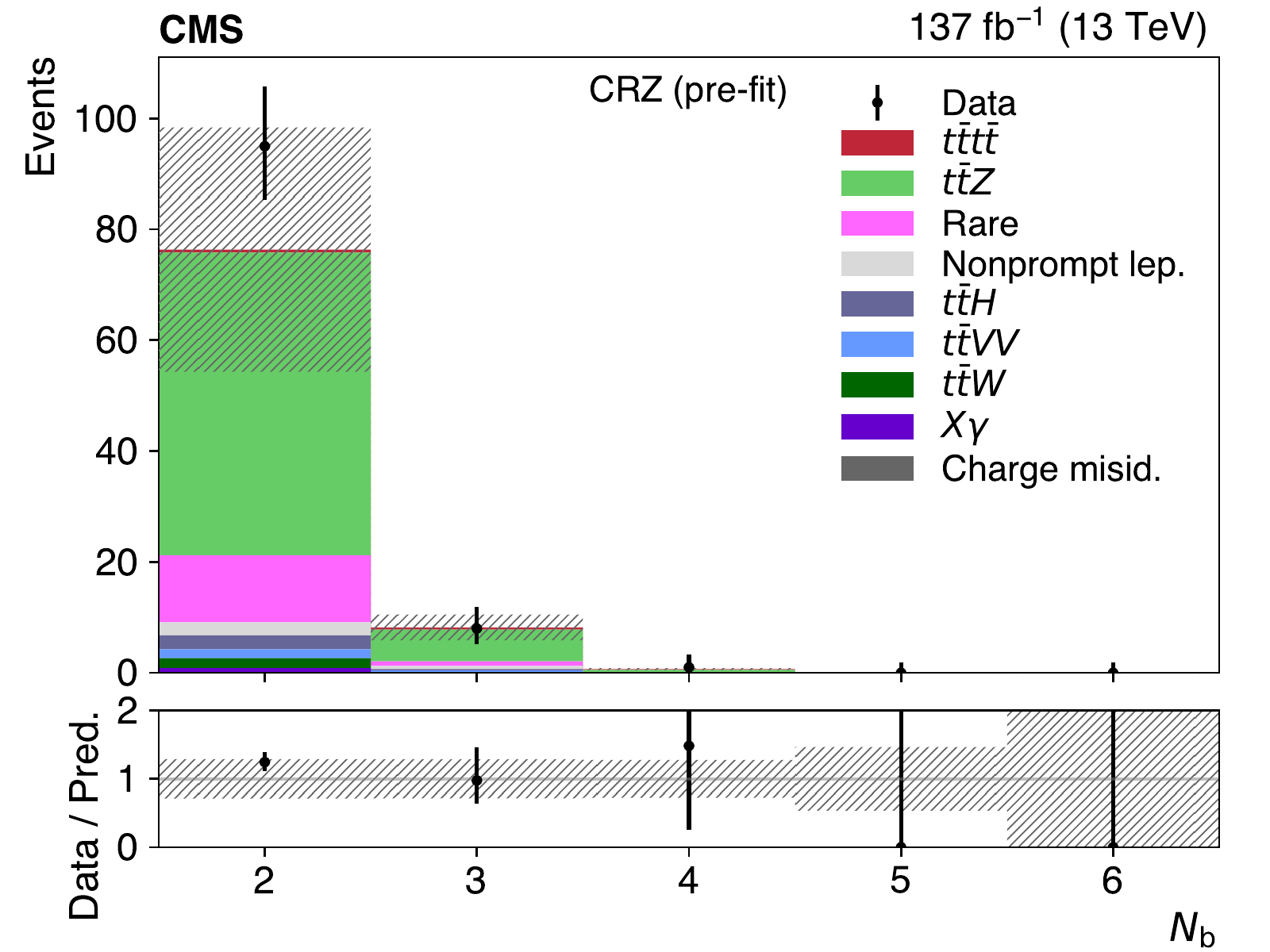}
    \caption{Distributions of \Njets (left) and \Nbjets (right) in the \ttW (upper) and \ttZ (lower) CRs, before fitting to data.
The hatched areas represent the  uncertainties in the SM signal and background predictions.
The \tttt signal assumes the SM cross section from Ref.~\cite{Frederix:2017wme}.
 The lower panels show the ratios of the observed event yield to the total prediction of
    signal plus background.
    }
\label{fig:kinemcr}
\end{figure*}

Distributions of the main kinematic variables (\Njets, \Nbjets, \HT, and \ptmiss) for events in the baseline region, as defined in Section~\ref{sec:selection},
are shown in Fig.~\ref{fig:kinemsr} and compared to the SM background predictions.
The \Njets and \Nbjets distributions for the CRW and CRZ are shown in Fig.~\ref{fig:kinemcr}.
The expected SM \tttt signal, normalized to its predicted cross section, is shown in both figures.
The SM predictions are statistically consistent with the observations.

A binned likelihood is constructed using the yields from the signal regions, the CRZ, as well as the CRW for the cut-based
analysis only, incorporating the experimental and theoretical uncertainties described in Section~\ref{sec:systematics} as ``nuisance'' parameters.
The measured cross section for \tttt and the significance of the observation relative to the background-only hypothesis
are obtained from a profile maximum-likelihood fit, in which the parameter of interest is \xsectttt and all nuisance parameters are profiled,
following the procedures described in Refs.~\cite{ATL-PHYS-PUB-2011-011,PDG}.
In addition, an upper limit at 95\% confidence level (\CL) is set on \xsectttt
using the modified frequentist \CLs criterion~\cite{Junk:1999kv,Read:2002hq},
with the profile likelihood ratio test statistic and asymptotic approximation~\cite{Cowan:2010js}.
We verified the consistency between the asymptotic and fully toy-based methods.
Alternatively,
by considering the SM, including the \tttt process with the SM cross section and uncertainty~\cite{Frederix:2017wme}, as the null hypothesis,
the fit provides cross section upper limits on
BSM processes with new scalar and pseudoscalar particles, as discussed in Section~\ref{sec:interpretations}.

\begin{figure}[!hbtp]
\centering
\includegraphics[width=\cmsFigWidth]{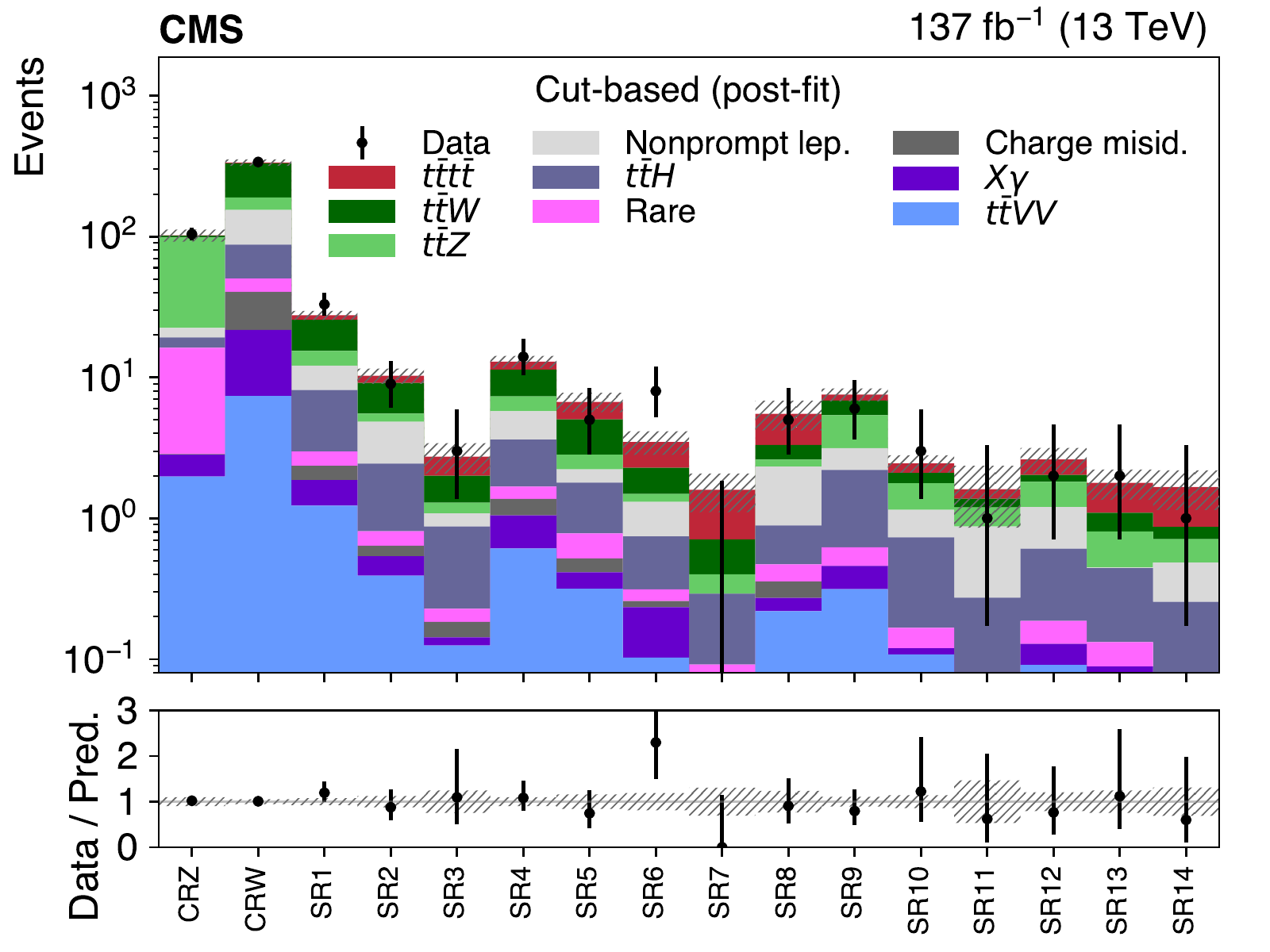} \\
\includegraphics[width=\cmsFigWidth]{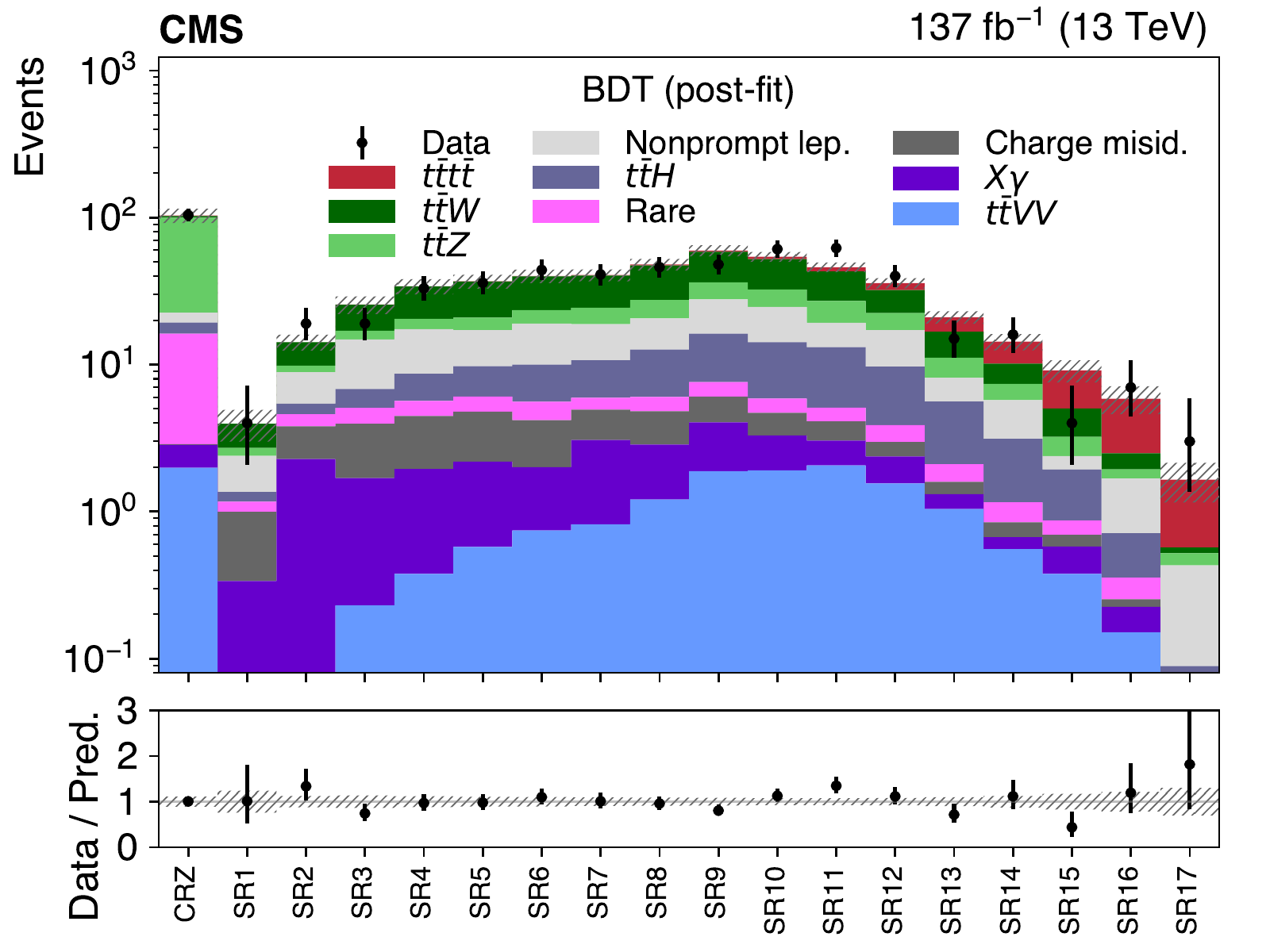}
\caption{ Observed yields in the control and signal regions for the cut-based (upper) and BDT (lower) analyses,
compared to the post-fit predictions for signal and background processes.
The hatched areas represent the total post-fit uncertainties in the signal and background predictions.
The lower panels show the ratios of the observed event yield to the total prediction of
    signal plus background.
    }
\label{fig:srcr}
\end{figure}

The values and uncertainties of most nuisance parameters are unchanged by the fit, but the ones significantly affected include those
corresponding to the \ttW and \ttZ normalizations, which are both scaled by $1.3\pm0.2$ by the fit,
in agreement with the ATLAS and CMS measurements of these processes~\cite{Aaboud:2019njj, Sirunyan:2017uzs, CMS-PAS-TOP-18-009}.
The predicted yields after the maximum-likelihood fit (post-fit) are compared to data in Fig.~\ref{fig:srcr} for the cut-based (upper) and BDT (lower) analyses,
where the fitted \tttt signal contribution is added to the background predictions.
The corresponding yields are shown in Tables~\ref{tab:srcryields} and \ref{tab:srdiscyields} for the cut-based and BDT analysis, respectively.

The \tttt cross section and the 68\% \CL interval
 is measured to be
$9.4^{+6.2}_{-5.6}\unit{fb}$
 in the cut-based analysis, and
$12.6^{+5.8}_{-5.2}\unit{fb}$ in the BDT analysis.
Relative to the background-only hypothesis, the observed and expected significances are 1.7 and 2.5
standard deviations, respectively, for the cut-based analysis, and  2.6 and 2.7 standard deviations for the BDT analysis.
The observed 95\% \CL upper limits on the cross section
are $20.0\unit{fb}$ in the cut-based and  $22.5\unit{fb}$ in the BDT analyses.
The corresponding expected upper limits on the \tttt cross section, assuming no SM \tttt contribution to the data, are $9.4^{+4.3}_{-2.9}\unit{fb}$ (cut-based)
and $8.5^{+3.9}_{-2.6}\unit{fb}$  (BDT), a significant improvement relative to the value of $20.8^{+11.2}_{-6.9}\unit{fb}$ of Ref.~\cite{tttt2016}.
The BDT and cut-based observed results were found to be statistically compatible by using correlated toy pseudo-data sets.
We consider the BDT analysis as the primary result of this paper, as it provides a
higher expected measurement precision, and use the results from it for further interpretations in the following section.

\begin{table*}[htb!]
\centering
    \topcaption{
        The post-fit predicted background, $\tttt$ signal, and total yields with
          their total uncertainties and the observed number of events
          in the control and signal regions in data for the cut-based analysis.
}
\label{tab:srcryields}
    \begin{tabular}{ccccc}
        & SM background  & $\tttt$   & Total   & Observed \\
            \hline

CRZ  & $101\pm10\ \ $   & $0.83\pm0.49$ & $102\pm10\ \ $   & 104 \\
CRW  & $331\pm19\ \ $   & $3.9\pm2.3$   & $335\pm18\ \ $   & 338 \\
SR1  & $25.6\pm2.1\ \ $ & $2.0\pm1.2$   & $27.6\pm2.1\ \ $ & 33 \\
SR2  & $9.1\pm1.3$      & $1.13\pm0.65$ & $10.3\pm1.3\ \ $ & 9 \\
SR3  & $2.01\pm0.58$    & $0.73\pm0.42$ & $2.74\pm0.67$    & 3 \\
SR4  & $11.3\pm1.3\ \ $ & $1.58\pm0.90$ & $12.9\pm1.3\ \ $ & 14 \\
SR5  & $5.03\pm0.77$    & $1.68\pm0.95$ & $6.7\pm1.1$      & 5 \\
SR6  & $2.29\pm0.40$    & $1.20\pm0.67$ & $3.48\pm0.66$    & 8 \\
SR7  & $0.71\pm0.20$    & $0.88\pm0.48$ & $1.59\pm0.49$    & 0 \\
SR8  & $3.31\pm0.95$    & $2.2\pm1.3$   & $5.5\pm1.3$      & 5 \\
SR9  & $6.84\pm0.80$    & $0.71\pm0.39$ & $7.55\pm0.80$    & 6 \\
SR10 & $2.10\pm0.31$    & $0.35\pm0.22$ & $2.45\pm0.35$    & 3 \\
SR11 & $1.38\pm0.75$    & $0.23\pm0.14$ & $1.61\pm0.75$    & 1 \\
SR12 & $2.03\pm0.48$    & $0.59\pm0.34$ & $2.62\pm0.54$    & 2 \\
SR13 & $1.09\pm0.28$    & $0.69\pm0.39$ & $1.78\pm0.44$    & 2 \\
SR14 & $0.87\pm0.30$    & $0.80\pm0.45$ & $1.67\pm0.52$    & 1 \\

\end{tabular}

\end{table*}

\begin{table*}[htb!]
\centering
    \topcaption{
        The post-fit predicted background and $\tttt$ signal, and total yields with
          their total uncertainties and the observed number of events
          in the control and signal regions in data for the BDT analysis.
}
\label{tab:srdiscyields}
    \begin{tabular}{ccccc}
        & SM background  & $\tttt$   & Total   & Observed \\
            \hline

CRZ  & $102\pm12\ \ $   & $1.11\pm0.43$ & $103\pm12\ \ $   & 104 \\
SR1  & $3.95\pm0.96$    & $ <0.01 $     & $3.96\pm0.96$    & 4 \\
SR2  & $14.2\pm1.8\ \ $ & $0.01\pm0.01$ & $14.2\pm1.8\ \ $ & 19 \\
SR3  & $25.5\pm3.5\ \ $ & $0.04\pm0.03$ & $25.6\pm3.5\ \ $ & 19 \\
SR4  & $34.0\pm4.0\ \ $ & $0.08\pm0.05$ & $34.0\pm4.0\ \ $ & 33 \\
SR5  & $36.7\pm4.0\ \ $ & $0.15\pm0.07$ & $36.8\pm4.0\ \ $ & 36 \\
SR6  & $39.8\pm4.2\ \ $ & $0.23\pm0.12$ & $40.0\pm4.2\ \ $ & 44 \\
SR7  & $40.3\pm3.7\ \ $ & $0.31\pm0.16$ & $40.6\pm3.8\ \ $ & 41 \\
SR8  & $47.3\pm4.3\ \ $ & $0.72\pm0.28$ & $48.0\pm4.3\ \ $ & 46 \\
SR9  & $58.5\pm5.2\ \ $ & $1.18\pm0.46$ & $59.7\pm5.2\ \ $ & 48 \\
SR10 & $52.1\pm4.3\ \ $ & $1.91\pm0.74$ & $54.1\pm4.2\ \ $ & 61 \\
SR11 & $43.0\pm3.5\ \ $ & $3.0\pm1.2$   & $46.0\pm3.5\ \ $ & 62 \\
SR12 & $32.1\pm3.0\ \ $ & $3.7\pm1.4$   & $35.8\pm2.9\ \ $ & 40 \\
SR13 & $16.7\pm1.6\ \ $ & $4.3\pm1.6$   & $21.0\pm2.0\ \ $ & 15 \\
SR14 & $10.1\pm1.2\ \ $ & $4.2\pm1.6$   & $14.3\pm1.8\ \ $ & 16 \\
SR15 & $5.03\pm0.77$    & $4.1\pm1.5$   & $9.1\pm1.6$      & 4 \\
SR16 & $2.49\pm0.61$    & $3.4\pm1.3$   & $5.9\pm1.3$      & 7 \\
SR17 & $0.57\pm0.36$    & $1.08\pm0.42$ & $1.65\pm0.50$    & 3 \\

\end{tabular}

\end{table*}

\section{Interpretations}
\label{sec:interpretations}

This analysis is used to constrain SM parameters, as well as production of BSM particles and operators that can affect the \tttt production rate.
The existence of \tttt Feynman diagrams with virtual Higgs bosons allows interpreting the upper limit on \xsectttt as a constraint on the Yukawa coupling, $y_{\PQt}$, between
the top quark and the Higgs boson~\cite{TopYukawaTTTT, TopYukawaTTTTnew}.
Similarly, the measurement can be interpreted as a constraint on the Higgs boson oblique parameter
$\hat{H}$, defined as the Wilson coefficient of the dimension-six BSM operator modifying the Higgs boson propagator~\cite{ObliqueHiggs2019}.
More generically, Feynman diagrams where the virtual Higgs boson is replaced by a virtual BSM scalar ($\phi$) or vector ($\cPZpr$) particle with mass smaller than twice the top quark mass
($m < 2m_\PQt$), are used to interpret the result as a constraint on the couplings of such new particles~\cite{Alvarez:2016nrz}.
In addition, new particles with $m > 2m_\PQt$, such as a heavy scalar (\PH) or pseudoscalar (\PSA), can be produced on-shell in association with top quarks.
They can subsequently decay into top quark pairs, generating final states with three or four top quarks.
Constraints on the production of such heavy particles can be interpreted in terms of 2HDM parameters~\cite{Dicus:1994bm,Craig:2015jba,Craig:2016ygr},
or in the framework of simplified models of dark matter~\cite{Boveia:2016mrp, Albert:2017onk}.

\begin{figure}[!hbtp]
\centering
\includegraphics[width=.49\textwidth]{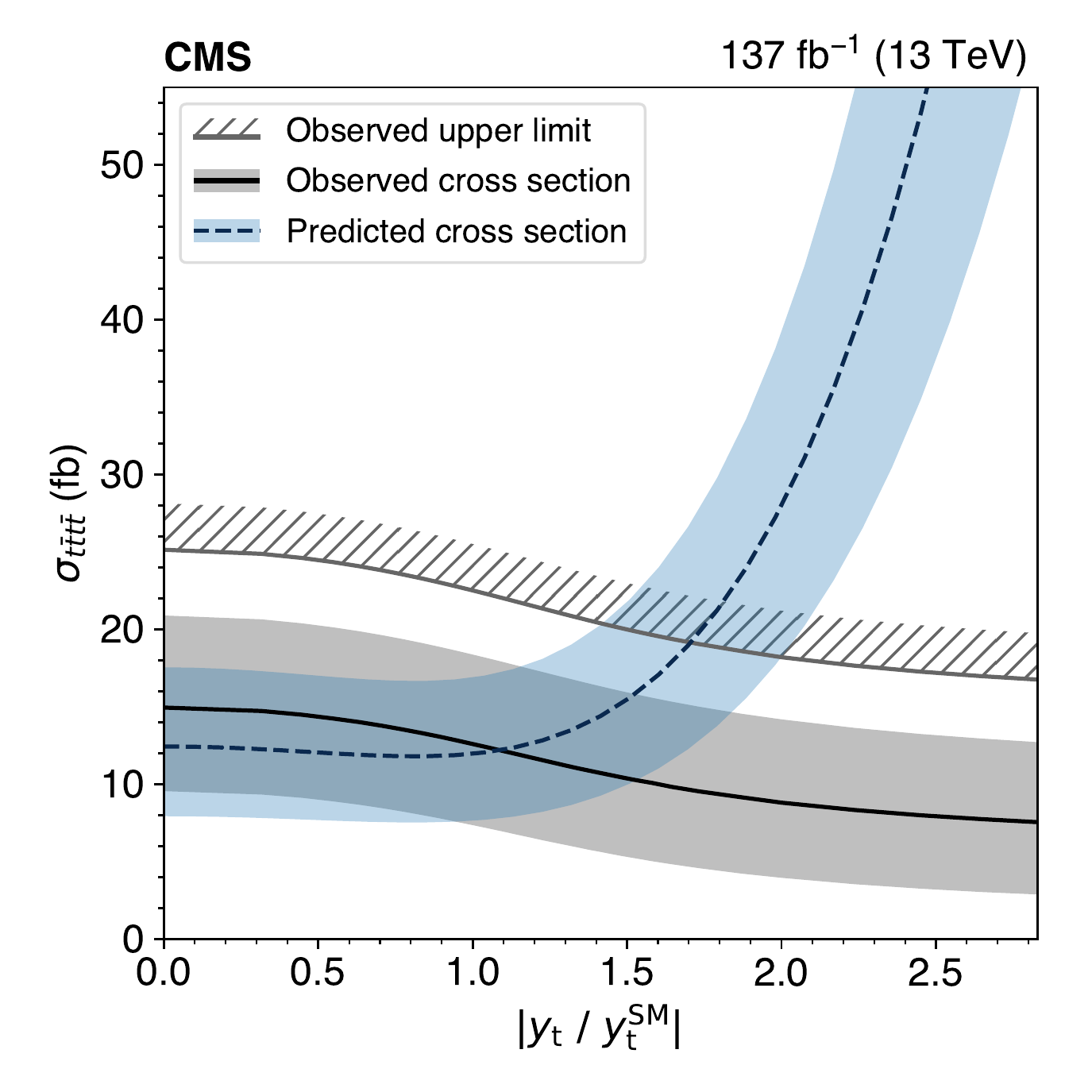}
\\
\caption{
    The observed \xsectttt (solid line) and 95\% \CL upper limit (hatched line) are shown as a function
    of $\abs{y_{\PQt}/y_{\PQt}^{\mathrm{SM}}}$. The predicted value (dashed line)~\cite{TopYukawaTTTT},
    calculated at LO and scaled to the calculation from Ref.~\cite{Frederix:2017wme}, is also plotted.
    The shaded band around the measured value gives the total uncertainty, while the shaded band around
    the predicted curve shows the theoretical uncertainty associated with the renormalization and
    factorization scales.
}
\label{fig:yukawa}
\end{figure}

When using our \tttt to determine a constraint on $y_{\PQt}$, we verified using a
LO simulation that the signal acceptance is not affected by the relative contribution of the virtual Higgs boson Feynman diagrams.
We take into account the dependence of the backgrounds on $y_{\PQt}$ by scaling the
\ttH cross section by $\abs{y_{\PQt}/y_{\PQt}^{\mathrm{SM}}}^2$ prior to the fit,  where $y_{\PQt}^{\mathrm{SM}}$ represents the SM value of the top quark Yukawa coupling.
As a result of the \ttH background rescaling, the measured \xsectttt depends on $\abs{y_{\PQt}/y_{\PQt}^{\mathrm{SM}}}$, as shown in Fig.~\ref{fig:yukawa}.
The measurement is compared to the theoretical prediction obtained from the LO calculation of Ref.~\cite{TopYukawaTTTT},
scaled to the $12.0^{+2.2}_{-2.5}\unit{fb}$ cross section obtained in Ref.~\cite{Frederix:2017wme},
and including the uncertainty associated with doubling and halving the renormalization and factorization scales.
Comparing the observed limit on \xsectttt with the central, upper, and lower values of its theoretical prediction, we obtain 95\% \CL limits of
$\abs{y_{\PQt}/y_{\PQt}^{\mathrm{SM}}} < 1.7$, $1.4$, and $2.0$, respectively,
an improvement over the previous CMS result~\cite{tttt2016}. Alternatively, assuming that the on-shell Yukawa coupling is equal to that of the SM,
we do not rescale the \ttH background with respect to its SM prediction, and obtain corresponding limits on the off-shell Yukawa coupling of
$\abs{y_{\PQt}/y_{\PQt}^{\mathrm{SM}}} < 1.8$, $1.5$, and $2.1$.
Since $y_{\PQt}$ affects the Higgs boson production cross section in both the
gluon fusion and \ttH modes, constraints on $y_{\PQt}$ can also be obtained from
a combination of Higgs boson measurements~\cite{AtlasCmsHiggsComb}.
However, these constraints require assumptions about the total width of the Higgs boson, while the \tttt-based limit does not.
For the $\hat{H}$ interpretation, the BDT analysis is repeated using simulated samples of \tttt signal events with different values
of $\hat{H}$ to account for small acceptance and kinematic differences, as described in Section~\ref{sec:samples}.
 We rescale the \ttH cross section by $(1-\hat{H})^2$ to account for its $\hat{H}$ dependency~\cite{ObliqueHiggs2019}.
This results in the 95\% \CL upper limit of $\hat{H} < 0.12$.
For reference, the authors of Ref.~\cite{ObliqueHiggs2019} used recent LHC on-shell Higgs boson measurements to set a constraint of $\hat{H} < 0.16$ at 95\% \CL.

\begin{figure}[!hbtp]
\centering
    \includegraphics[width=.49\textwidth]{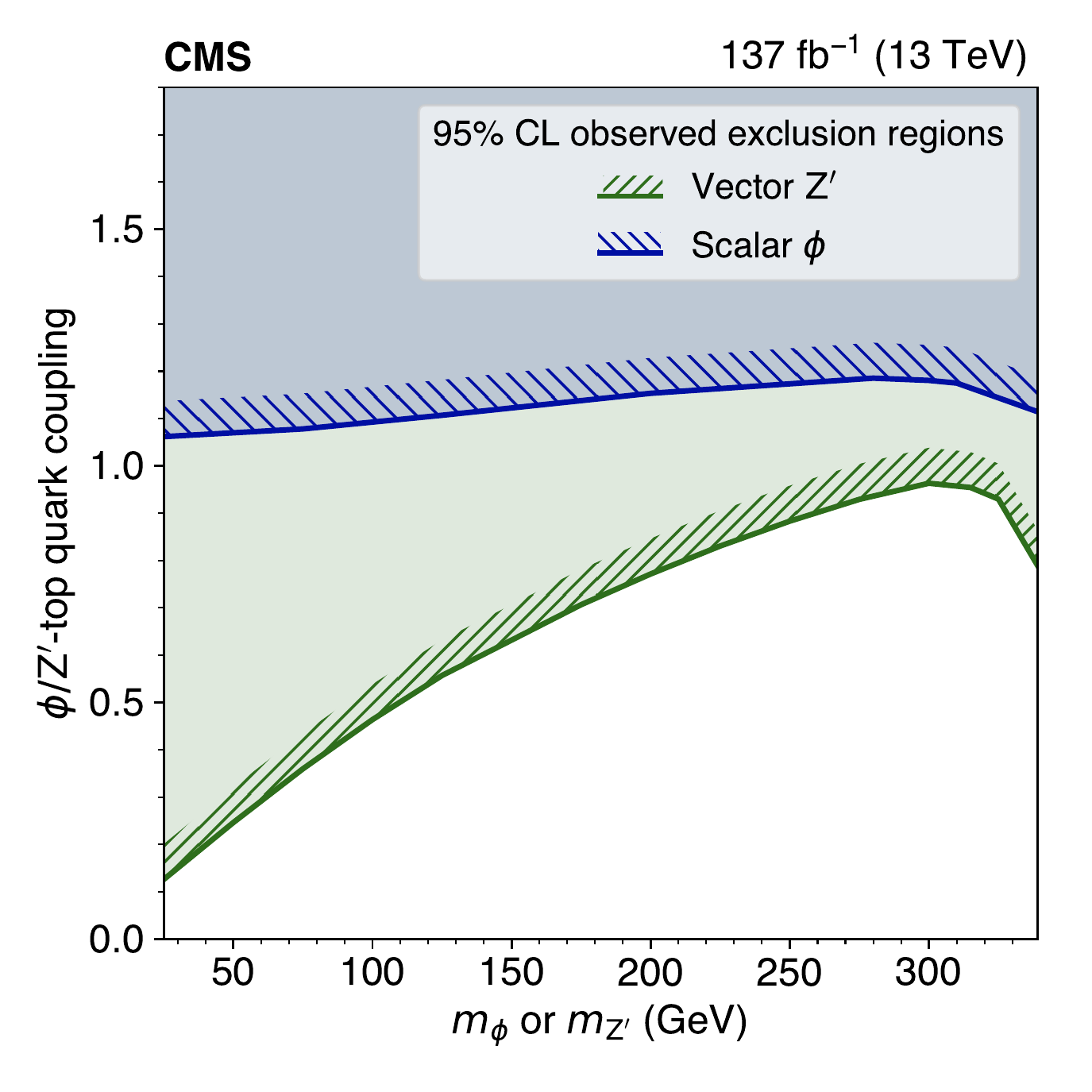}
\caption{
    The 95\% \CL exclusion regions in the plane of the $\phi/\cPZpr$-top quark coupling versus
    $m_{\phi}$ or $m_{\cPZpr}$. The excluded regions are above the hatched lines.
    }
\label{fig:ZprimePhiExclusions}
\end{figure}

To study the off-shell effect of new particles with $m < 2m_\PQt$, we first consider neutral scalar ($\phi$) and neutral vector ($\cPZpr$) particles
that couple to top quarks.
Such particles are at present only weakly constrained, while they can give significant contributions to the \tttt cross section~\cite{Alvarez:2016nrz}.
Having verified in LO simulation that these new particles affect the signal acceptance by less than 10\%,
we recalculate the \xsectttt upper limit of the BDT analysis including an additional 10\% uncertainty in the acceptance,
and obtain the 95\% \CL upper limit of 23.0\unit{fb} on the total \tttt cross section, slightly weaker than the 22.5\unit{fb} limit obtained in Section~\ref{sec:results}.
Comparing this upper limit to the predicted cross section in models where \tttt production includes a $\phi$ or a $\cPZpr$ in addition to SM contributions and associated interference, we set limits on the masses
and couplings of these new particles, shown in Fig.~\ref{fig:ZprimePhiExclusions}.
These limits exclude couplings larger than 1.2 for $m_{\phi}$ in the 25--340\GeV range and larger than 0.1 (0.9) for $m_{\cPZpr} = 25$ (300)\GeV.

\begin{figure*}[!hbtp]
\centering
    \includegraphics[width=.49\textwidth]{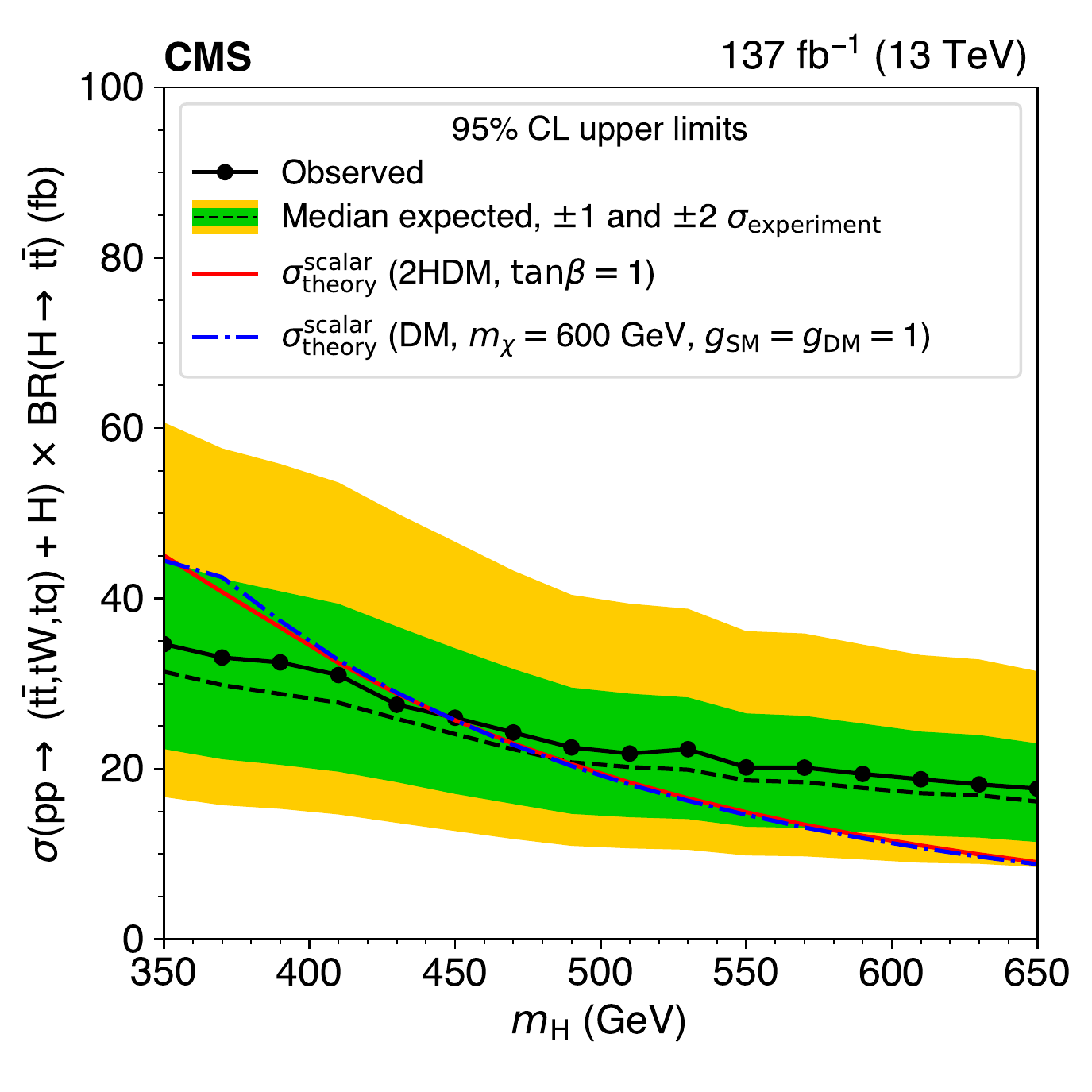}
    \includegraphics[width=.49\textwidth]{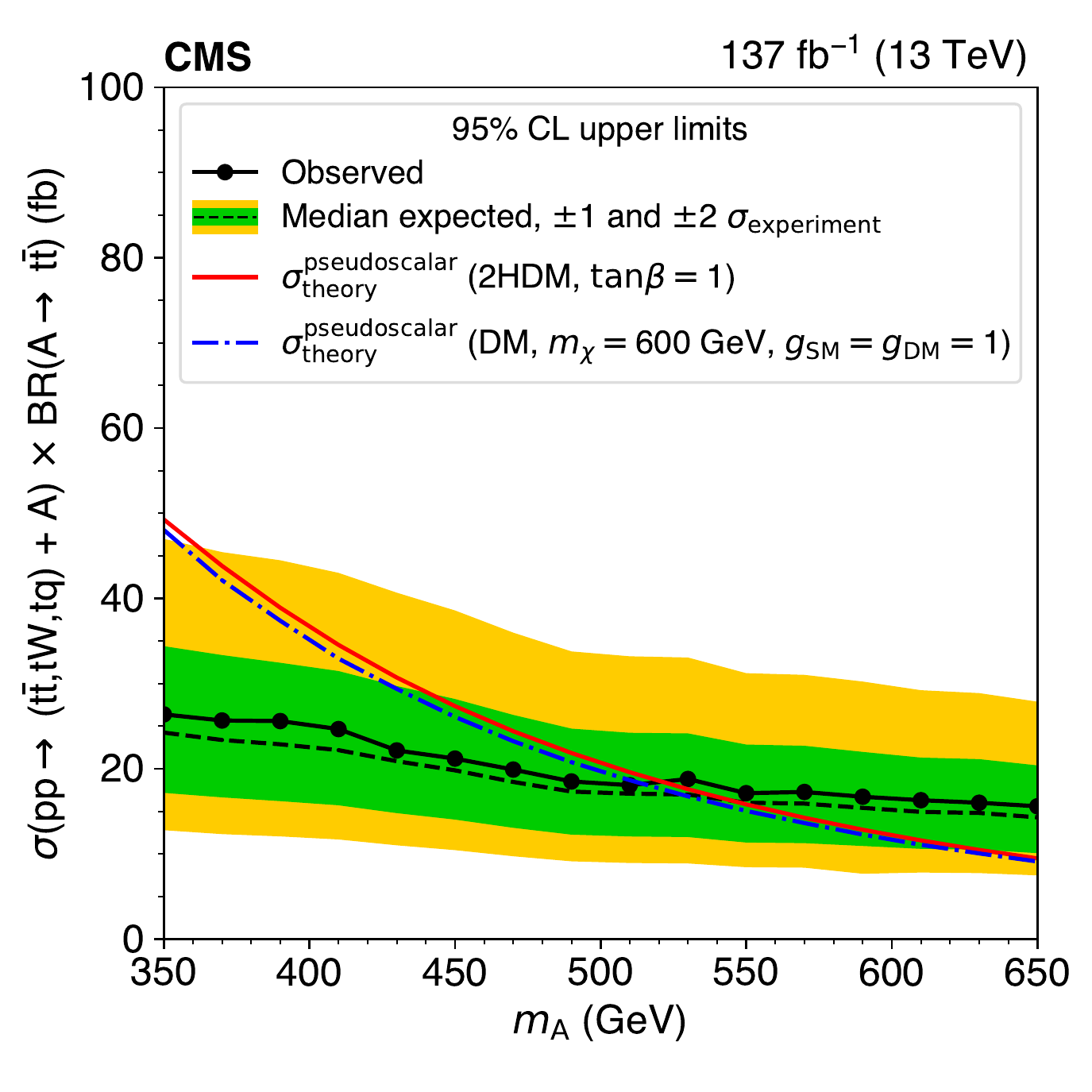}
\caption{
    The observed (points) and expected (dashed line) 95\% \CL upper limits on the cross section
    times branching fraction to \ttbar for the production of a new heavy scalar \PH (left) and pseudoscalar \PSA (right),
    as a function of mass. The inner and outer bands around the expected limits indicate the regions containing 68 and 95\%,
    respectively, of the distribution of limits under the background-only hypothesis. Theoretical values are shown for Type-II 2HDM
    in the alignment limit (solid line) and simplified dark matter (dot-dashed line) models.
}
\label{fig:HiggsLimits}
\end{figure*}

\begin{figure*}[!hbtp]
\centering
    \includegraphics[width=.49\textwidth]{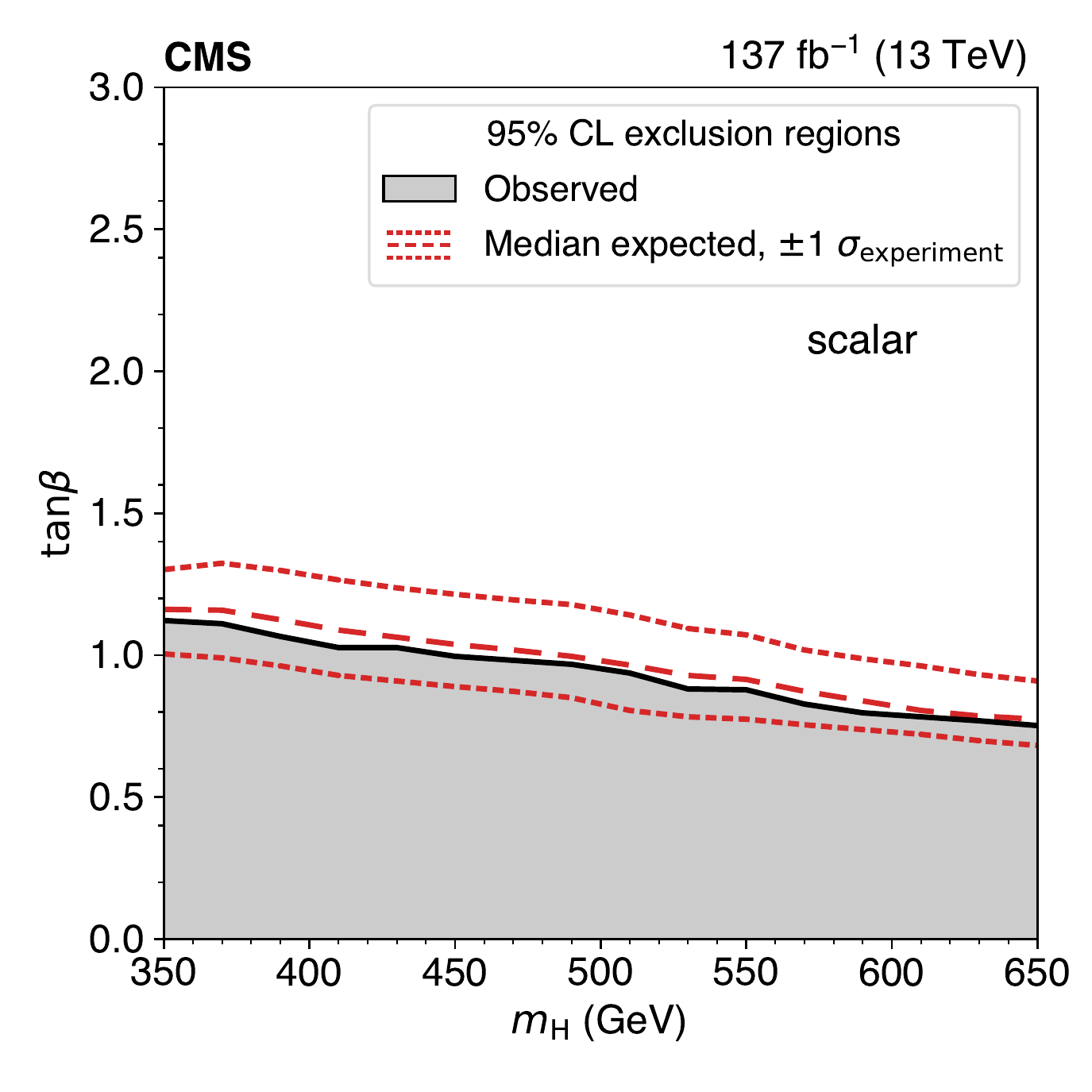}
    \includegraphics[width=.49\textwidth]{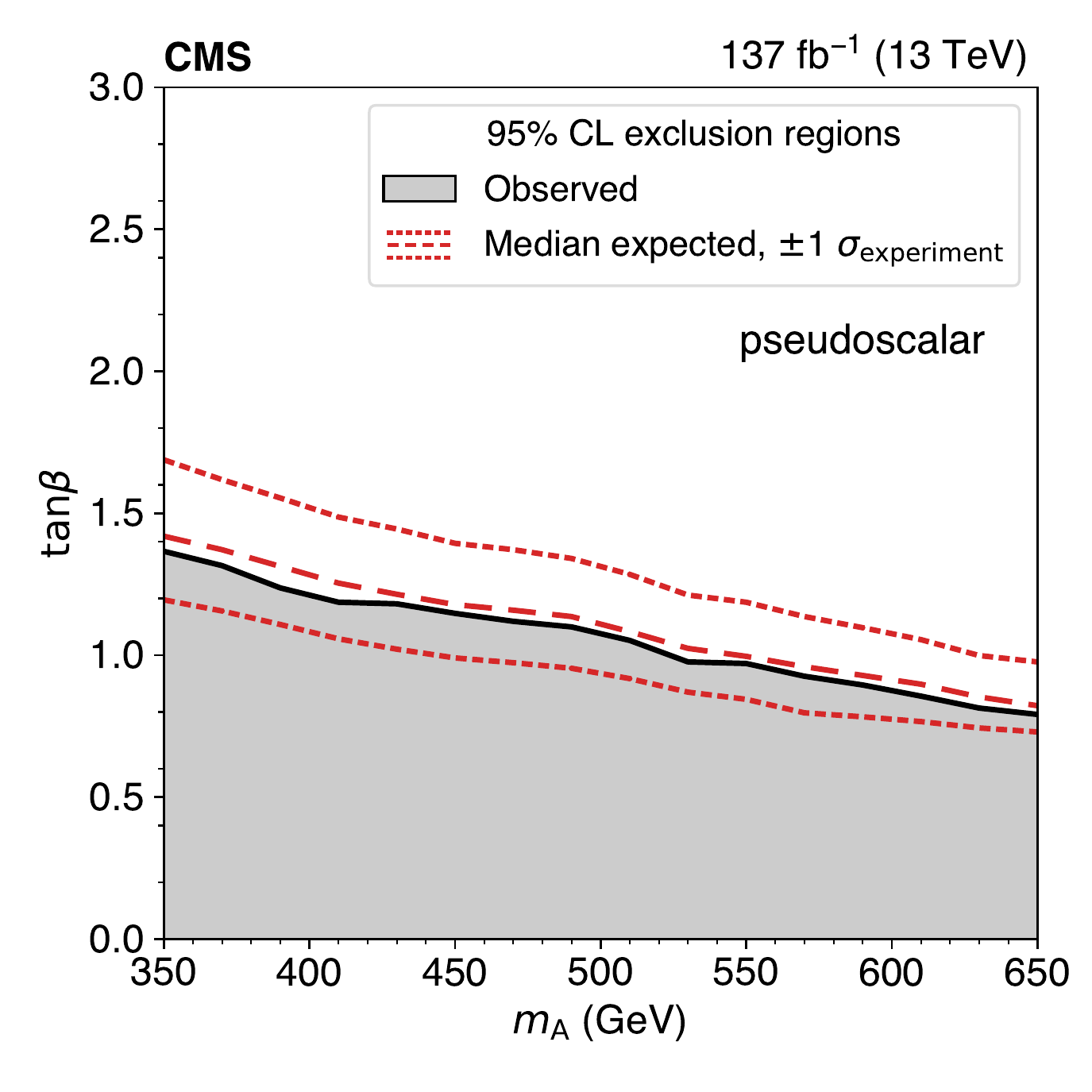}
    \includegraphics[width=.49\textwidth]{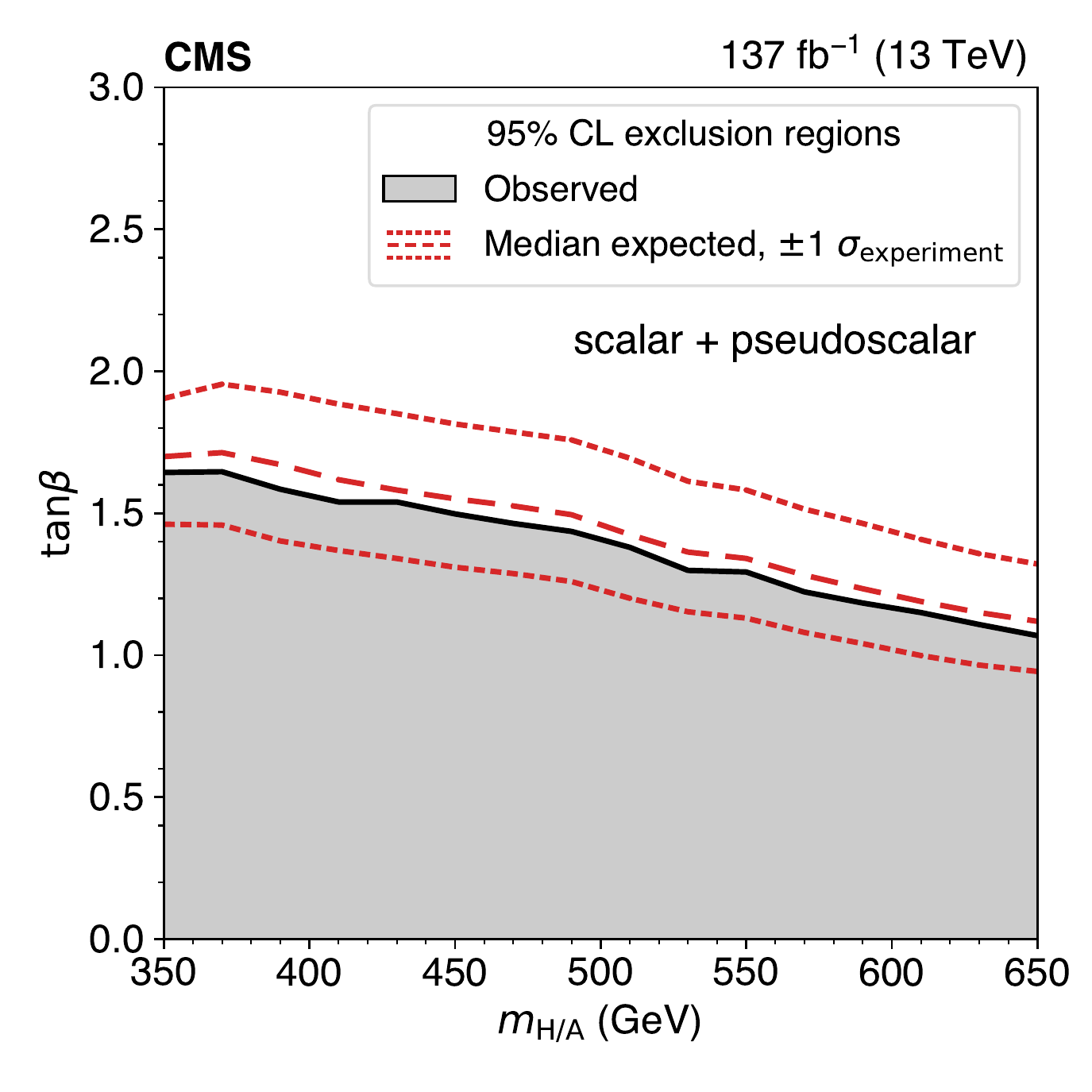}
\caption{
    The observed (solid curve) and expected (long-dashed curve) 95\% \CL exclusion regions in the $\tan\beta$ versus mass plane
    for Type-II 2HDM models in the alignment limit for a new scalar \PH (upper left), pseudoscalar \PSA (upper right), and both (lower) particles.
    The short-dashed curves around the expected limits indicate the region containing 68\% of the distribution of limits expected under the
    background-only hypothesis. The excluded regions are below the curves.
}
\label{fig:HiggsLimitsTB}
\end{figure*}

\begin{figure*}[!hbtp]
\centering
    \includegraphics[width=.49\textwidth]{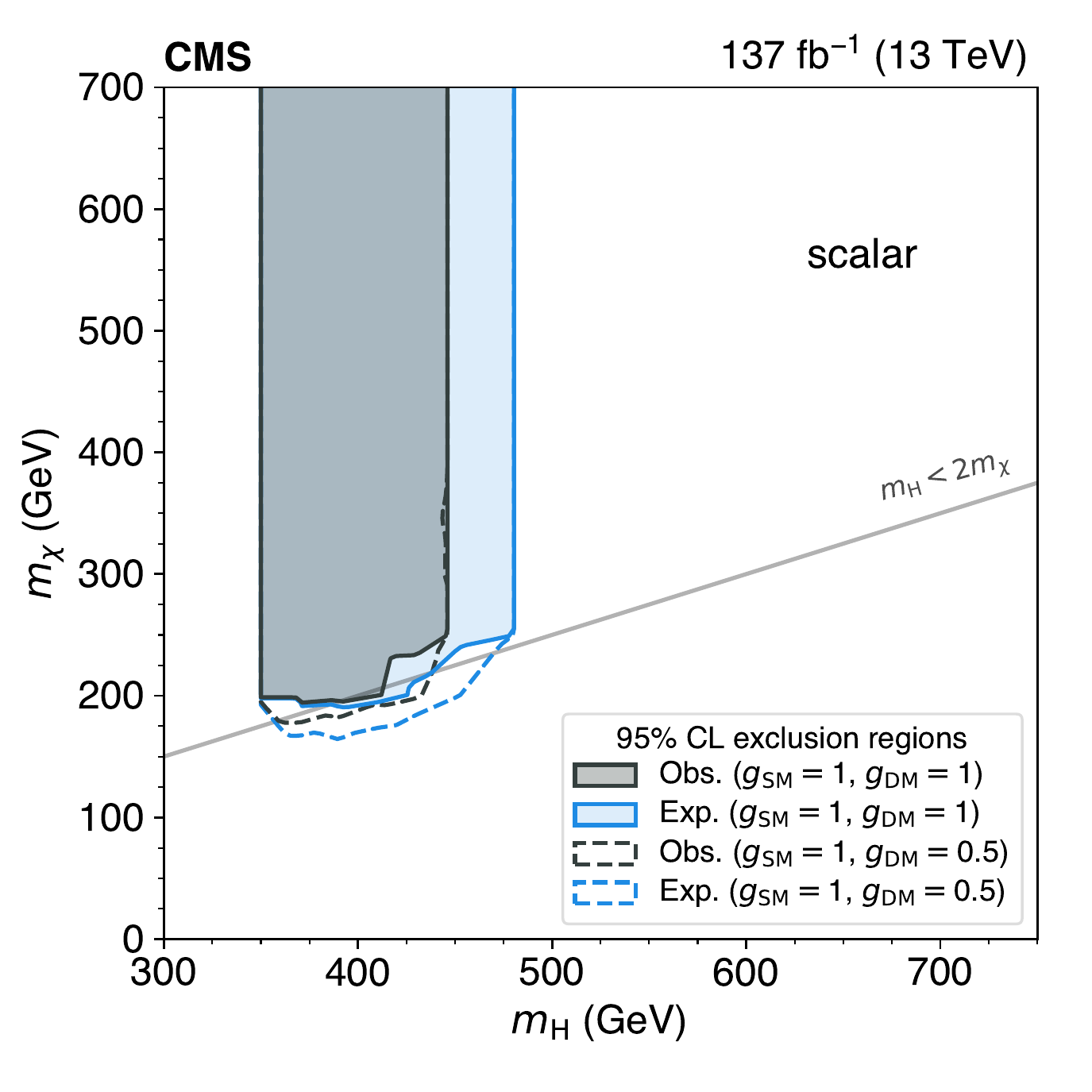}
    \includegraphics[width=.49\textwidth]{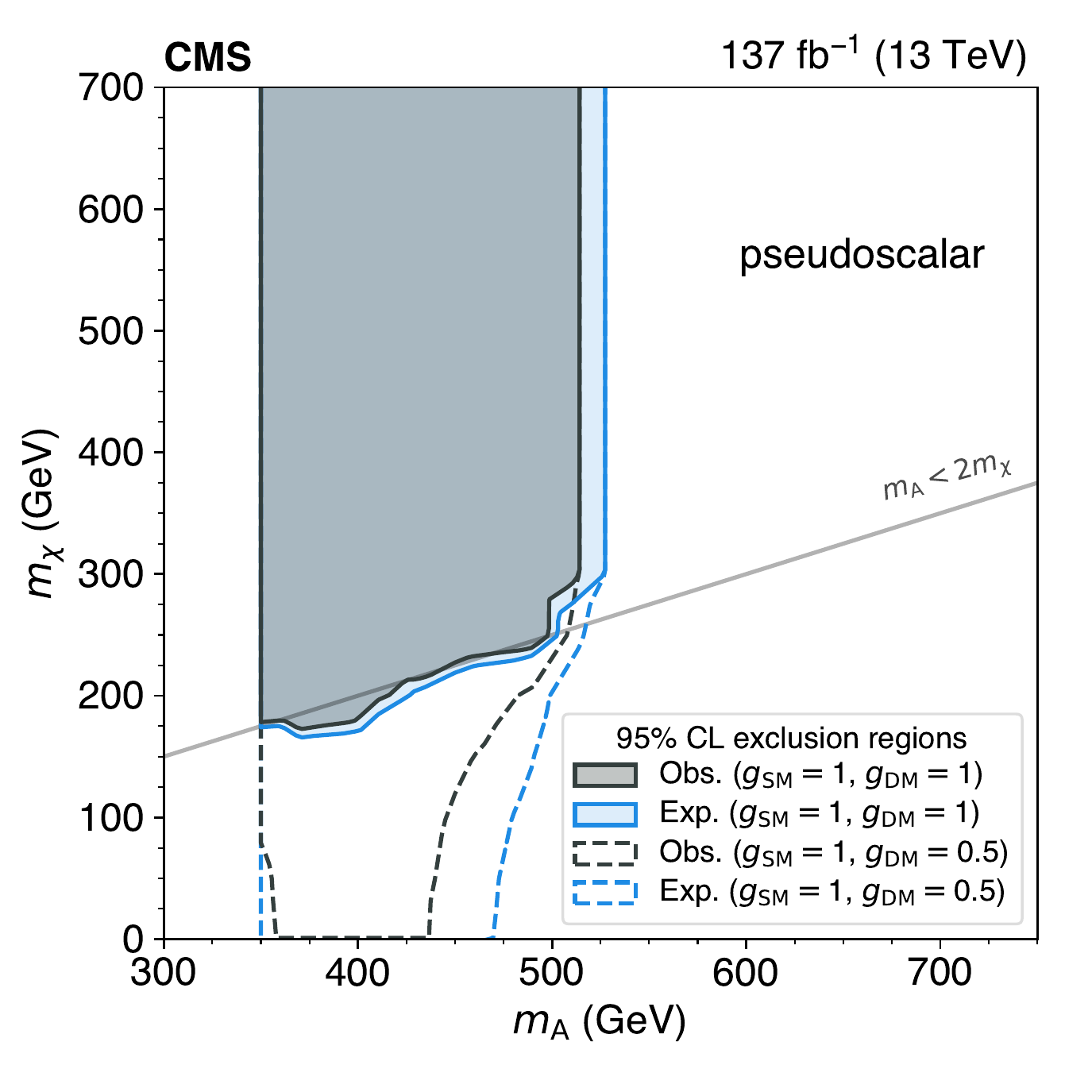}
\caption{
	Exclusion regions at 95\% \CL in the plane of $m_\chi$ vs. $m_{\PH}$ (left) or $m_{\PSA}$ (right).
    The outer lighter and inner darker solid curves show the expected and observed limits, respectively,
    assuming $g_\mathrm{SM} = g_\mathrm{DM} = 1$. The excluded regions, shaded, are above the limit curves.
    The dashed lines show the limits assuming a weaker coupling between $\PH/\PSA$ and $\chi$, $g_\mathrm{DM} = 0.5$.
    }
\label{fig:DMLimits}
\end{figure*}

We consider on-shell effects from new scalar and pseudoscalar particles with $m > 2m_\PQt$. At such masses, the production rate of these particles
in association with a single top quark ($\PQt\PQq\PH/\PSA$, $\PQt\PW\PH/\PSA$) becomes significant,
so we include these processes in addition to $\PQt\overline{\PQt}\PH/\PSA$.
As pointed out in Ref.~\cite{Craig:2016ygr}, these processes do not suffer significant interference with the SM \tttt process.
To obtain upper limits on the sum of these processes followed by the decay $\PH/\PSA\to \ttbar$,
we use the BDT analysis and treat the SM \tttt process as a background.
Figure~\ref{fig:HiggsLimits} shows the excluded cross section as a function of the mass of the scalar (left) and pseudoscalar (right).
Comparing these limits with the Type-II 2HDM cross sections with $\tan\beta = 1$ in the alignment limit,
we exclude scalar (pseudoscalar) masses up to 470~(550)\GeV, improving by more than 100\GeV with respect to the previous CMS limits~\cite{SS2016}.
Alternatively, we consider the simplified model of dark matter defined in Ref.~\cite{DMsingletopCMS},
which includes a Dirac fermion dark matter candidate, $\chi$, in addition to $\PH/\PSA$, and where the couplings of $\PH/\PSA$
to SM fermions and $\chi$ are determined by parameters $g_\mathrm{SM}$ and $g_\mathrm{DM}$, respectively.
In this model, exclusions similar to those from 2HDM are reached by assuming $g_\mathrm{SM} = 1$ and $g_\mathrm{DM} = 1$,
and taking  $m_{\PH/\PSA} < 2 m_\chi$.
Relaxing the 2HDM assumption of $\tan\beta = 1$, Fig.~\ref{fig:HiggsLimitsTB} shows the 2HDM limit as a function of $\PH/\PSA$ mass and $\tan\beta$,
considering one new particle at a time and also including a scenario with  $m_\PH = m_\PSA$ inspired by
a special case of Type-II 2HDM, the hMSSM~\cite{Djouadi:2013uqa}.
Values of $\tan\beta$ up to 0.8--1.6 are excluded, depending on the assumptions made.
These exclusions are comparable to those of a recent CMS search for the resonant production of $\PH/\PSA$ in the
$\Pp\to \PH/\PSA \to \ttbar$ channel~\cite{HIG-17-027}.
Relaxing the $m_{\PH/\PSA} < 2 m_\chi$ assumption in the dark matter model, Fig.~\ref{fig:DMLimits} shows the
limit in this model as a function of the masses of both $\PH/\PSA$ and $\chi$,
for $g_\mathrm{DM} = 1$ and for two different assumptions of $g_\mathrm{SM}$.
Large sections of the phase space of simplified dark matter models are excluded, and the reach of this analysis is complementary
to that of analyses
considering decays of $\PH/\PSA$ into invisible dark matter candidates, such as those of Refs.~\cite{DMsingletopCMS, Aaboud:2017rzf}.

\section{Summary}
\label{sec:summary}

 The standard model (SM) production of \tttt  has been studied in data from
 $\sqrt{s} = 13\TeV$ proton-proton collisions collected using the CMS detector during the LHC 2016--2018 data-taking period,
 corresponding to an integrated luminosity of \sslumi.
The final state with either two same-sign leptons or at least three leptons is analyzed using two  strategies,
 the first relying on a cut-based categorization in lepton and jet multiplicity and jet flavor,
 the second taking advantage of a multivariate approach to distinguish the \tttt signal from its many backgrounds.
 The more precise multivariate strategy yields an observed (expected) significance of 2.6 (2.7)
standard deviations relative to the background-only hypothesis,
 and a measured value for the \tttt cross section of
$12.6^{+5.8}_{-5.2}\unit{fb}$.
The results based on the two strategies are
in agreement with each other and with the SM prediction of $12.0^{+2.2}_{-2.5}\unit{fb}$~\cite{Frederix:2017wme}.

The results of the boosted decision tree (BDT) analysis are also used to constrain
the top quark Yukawa coupling $y_{\PQt}$ relative to its SM value,
based on the $\abs{y_{\PQt}}$ dependence of \xsectttt calculated at leading order in Ref.~\cite{TopYukawaTTTT},
resulting in the 95\% confidence level (\CL) limit of
 $\abs{y_{\PQt}/y_{\PQt}^{\mathrm{SM}}} < 1.7$.
The Higgs boson oblique parameter in the effective field theory framework~\cite{ObliqueHiggs2019} is similarly constrained to $\hat{H} < 0.12$ at 95\% \CL.
Upper limits ranging from 0.1 to 1.2 are also set on the coupling between the top quark and a new scalar ($\phi$) or vector ($\cPZpr$) particle with mass less than twice that of
the top quark ($m_\PQt$)~\cite{Alvarez:2016nrz}.
For new scalar ($\PH$) or pseudoscalar ($\PSA$) particles with $m > 2m_\PQt$, and decaying to \ttbar, their production in association with a single top quark or a
 top quark pair is probed.
The resulting cross section upper limit, between 15 and 35\unit{fb} at 95\% \CL, is interpreted in the context of Type-II two-Higgs-doublet
models~\cite{Dicus:1994bm,Craig:2015jba,Craig:2016ygr,Djouadi:2013uqa}
as a function of $\tan \beta$ and $m_\mathrm{\PH/\PSA}$, and in the context of simplified dark matter models~\cite{Boveia:2016mrp, Albert:2017onk},
as a function of $m_\mathrm{\PH/\PSA}$ and the mass of the dark matter candidate.

\begin{acknowledgments}
We congratulate our colleagues in the CERN accelerator departments for the excellent performance of the LHC and thank the technical and administrative staffs at CERN and at other CMS institutes for their contributions to the success of the CMS effort. In addition, we gratefully acknowledge the computing centers and personnel of the Worldwide LHC Computing Grid for delivering so effectively the computing infrastructure essential to our analyses. Finally, we acknowledge the enduring support for the construction and operation of the LHC and the CMS detector provided by the following funding agencies: BMBWF and FWF (Austria); FNRS and FWO (Belgium); CNPq, CAPES, FAPERJ, FAPERGS, and FAPESP (Brazil); MES (Bulgaria); CERN; CAS, MoST, and NSFC (China); COLCIENCIAS (Colombia); MSES and CSF (Croatia); RPF (Cyprus); SENESCYT (Ecuador); MoER, ERC IUT, PUT and ERDF (Estonia); Academy of Finland, MEC, and HIP (Finland); CEA and CNRS/IN2P3 (France); BMBF, DFG, and HGF (Germany); GSRT (Greece); NKFIA (Hungary); DAE and DST (India); IPM (Iran); SFI (Ireland); INFN (Italy); MSIP and NRF (Republic of Korea); MES (Latvia); LAS (Lithuania); MOE and UM (Malaysia); BUAP, CINVESTAV, CONACYT, LNS, SEP, and UASLP-FAI (Mexico); MOS (Montenegro); MBIE (New Zealand); PAEC (Pakistan); MSHE and NSC (Poland); FCT (Portugal); JINR (Dubna); MON, RosAtom, RAS, RFBR, and NRC KI (Russia); MESTD (Serbia); SEIDI, CPAN, PCTI, and FEDER (Spain); MOSTR (Sri Lanka); Swiss Funding Agencies (Switzerland); MST (Taipei); ThEPCenter, IPST, STAR, and NSTDA (Thailand); TUBITAK and TAEK (Turkey); NASU and SFFR (Ukraine); STFC (United Kingdom); DOE and NSF (USA).

\hyphenation{Rachada-pisek} Individuals have received support from the Marie-Curie program and the European Research Council and Horizon 2020 Grant, contract Nos.\ 675440, 752730, and 765710 (European Union); the Leventis Foundation; the A.P.\ Sloan Foundation; the Alexander von Humboldt Foundation; the Belgian Federal Science Policy Office; the Fonds pour la Formation \`a la Recherche dans l'Industrie et dans l'Agriculture (FRIA-Belgium); the Agentschap voor Innovatie door Wetenschap en Technologie (IWT-Belgium); the F.R.S.-FNRS and FWO (Belgium) under the ``Excellence of Science -- EOS" -- be.h project n.\ 30820817; the Beijing Municipal Science \& Technology Commission, No. Z181100004218003; the Ministry of Education, Youth and Sports (MEYS) of the Czech Republic; the Lend\"ulet (``Momentum") Program and the J\'anos Bolyai Research Scholarship of the Hungarian Academy of Sciences, the New National Excellence Program \'UNKP, the NKFIA research grants 123842, 123959, 124845, 124850, 125105, 128713, 128786, and 129058 (Hungary); the Council of Science and Industrial Research, India; the HOMING PLUS program of the Foundation for Polish Science, cofinanced from European Union, Regional Development Fund, the Mobility Plus program of the Ministry of Science and Higher Education, the National Science Center (Poland), contracts Harmonia 2014/14/M/ST2/00428, Opus 2014/13/B/ST2/02543, 2014/15/B/ST2/03998, and 2015/19/B/ST2/02861, Sonata-bis 2012/07/E/ST2/01406; the National Priorities Research Program by Qatar National Research Fund; the Ministry of Science and Education, grant no. 3.2989.2017 (Russia); the Programa Estatal de Fomento de la Investigaci{\'o}n Cient{\'i}fica y T{\'e}cnica de Excelencia Mar\'{\i}a de Maeztu, grant MDM-2015-0509 and the Programa Severo Ochoa del Principado de Asturias; the Thalis and Aristeia programs cofinanced by EU-ESF and the Greek NSRF; the Rachadapisek Sompot Fund for Postdoctoral Fellowship, Chulalongkorn University and the Chulalongkorn Academic into Its 2nd Century Project Advancement Project (Thailand); the Welch Foundation, contract C-1845; and the Weston Havens Foundation (USA).
\end{acknowledgments}

\ifthenelse{\boolean{cms@external}}{\clearpage}{}
\bibliography{auto_generated}
\cleardoublepage \appendix\section{The CMS Collaboration \label{app:collab}}\begin{sloppypar}\hyphenpenalty=5000\widowpenalty=500\clubpenalty=5000\vskip\cmsinstskip
\textbf{Yerevan Physics Institute, Yerevan, Armenia}\\*[0pt]
A.M.~Sirunyan$^{\textrm{\dag}}$, A.~Tumasyan
\vskip\cmsinstskip
\textbf{Institut f\"{u}r Hochenergiephysik, Wien, Austria}\\*[0pt]
W.~Adam, F.~Ambrogi, T.~Bergauer, J.~Brandstetter, M.~Dragicevic, J.~Er\"{o}, A.~Escalante~Del~Valle, M.~Flechl, R.~Fr\"{u}hwirth\cmsAuthorMark{1}, M.~Jeitler\cmsAuthorMark{1}, N.~Krammer, I.~Kr\"{a}tschmer, D.~Liko, T.~Madlener, I.~Mikulec, N.~Rad, J.~Schieck\cmsAuthorMark{1}, R.~Sch\"{o}fbeck, M.~Spanring, D.~Spitzbart, W.~Waltenberger, C.-E.~Wulz\cmsAuthorMark{1}, M.~Zarucki
% ...
\vskip\cmsinstskip
\dag: Deceased\\
1:  Also at Vienna University of Technology, Vienna, Austria\\
% ...
\end{sloppypar}
\end{document}